\documentclass[1p,numbered]{article}
\usepackage{graphicx}
\usepackage{natbib}
\usepackage{caption}
\usepackage{subcaption}
\usepackage[mathlines]{lineno}
\usepackage{hyperref}
\usepackage{setspace}
\usepackage{mathrsfs}
\usepackage{rotating}
\usepackage{url}
\usepackage{amssymb}
\usepackage{multirow}
\usepackage{comment}
\usepackage{color}
\usepackage{enumerate}
\DeclareGraphicsExtensions{.pdf,.png,.jpg}
\begin{document}
\begin{titlepage}

\thispagestyle{empty}
\title{sPEGG: high throughput eco-evolutionary simulations on commodity graphics processors}
\author{Kenichi W. Okamoto\dag \ddag \S $\sharp$ { and} Priyanga Amarasekare \dag $\natural$ \\ \dag \textit{Department of Ecology and Evolutionary Biology} \\ \textit{University of California, Los Angeles}\\ 
\ddag kenichi.okamoto@yale.edu\\
\S \textit{Department of Entomology, North Carolina State University}\\
$\sharp$ Present affiliation: \textit{Yale Institute for Biospheric Studies, Yale University} \\
$\natural$ amarasek@ucla.edu\\
Corresponding Author: \\
Kenichi W. Okamoto \\ 
Yale University \\ 
165 Prospect Street \\
New Haven CT 06511 USA \\
}

\maketitle
\thispagestyle{empty}
\end{titlepage}
\begin{abstract}
Integrating population genetics into community ecology theory is a major goal in ecology and evolution, but analyzing the resulting models is computationally daunting. Here we describe sPEGG (\underline{s}imulating \underline{P}henotypic \underline{E}volution on \underline{G}eneral Purpose \underline{G}raphics Processing Units (GPGPUs)), an open-source, multi-species forward-time population genetics simulator. Using a single commodity GPGPU instead of a single central processor, we find sPEGG can accelerate eco-evolutionary simulations by a factor of over 200, comparable to performance on a small-to-medium sized computer cluster. 

\end{abstract}
keywords: forward-time population genetics models; computational biology; eco-evolutionary dynamics; parallel computing; GPU; individual-based models; CUDA

\pagestyle{plain}

\section*{Main}
Elucidating how changes in gene frequencies within species drive ecological dynamics, and how ecological dynamics, in turn, affect gene frequencies is a central goal in ecology and evolutionary biology (\citealt{yoshida03, post09, schoener11}, Hendry \textit{in press}\nocite{hendry16}). Such eco-evolutionary feedbacks can potentially be quite complex, especially when the important ecological processes (for instance, inter- and intra-specific interactions) are underlain by quantitative traits influenced by several loci.  Purely phenotypic models that do not describe the underlying genetic details (e.g., classical quantitative genetics) have greatly enhanced our understanding of evolutionary processes. Yet rapidly growing knowledge of the genetic architecture of ecologically important traits (e.g., \citealt{stapley10, hendry13}) necessitates incorporating such genetic details into our models to advance our understanding of eco-evolutionary dynamics. Such models are needed to guide developing effective responses to anthropogenic perturbations of natural communities, improve agricultural practices and manage global health risks such as emergent pathogens.

Due to their inherent complexity, models coupling genetic and ecological dynamics must be analyzed through numerical simulation, often using individual-based models (IBMs) that explicitly characterize the fates of individuals and the alleles they carry (e.g., \citealt{deangelis05} and \citealt{carvajal10}). IBMs are attractive for several reasons. First, the same processes that drive changes in allele frequencies in nature (e.g., selection, recombination and mutation acting through reproduction, survivorship, dispersal, etc.) have a one-to-one correspondence in the model (e.g., \citealt{hoban12}). Second, their inherent stochasticity allows us to rigorously conduct statistical inference to test theory with data using likelihood functions based on biological principles, rather than mathematical convenience (e.g., \citealt{hartig11}). Finally, parameterizing IBMs is often easier than parameterizing alternative population models (e.g., \citealt{pacala96}). Thus, IBMs provide both a mechanistic explanation for the observed patterns of individual phenotypic variation, as well as a framework for forecasting how selection, drift and gene flow and constraints (e.g., pleiotropy; also, \citealt{futuyma10}) can drive evolutionary change. 

However, some of these advantages make using IBMs computationally challenging. Because they model realistic properties (e.g., age or size structure, species interactions) over evolutionary time scales (e.g., large number of generations), an individual run of a particular simulation can take a very long time to complete (\citealt{deangelis05}). Moreover, stochastic IBMs require several replicate simulations per parameter combination, and that several such combinations be analyzed. 

These performance issues constitute non-trivial concerns for investigators who face constraints on their time and/or funding. Modern central processing units (CPUs), the most commonly used type of processor by most biologists, are unlikely to see meaningful performance improvements in the near future (\citealt{fuller11}). This means IBMs need to be simulated simultaneously on multiple computers, which often requires considerable financial investment and institutional support.

Here we introduce an open-source software framework, sPEGG (\underline{s}imulating \underline{P}henotypic \underline{E}volution on \underline{G}eneral Purpose \underline{G}raphics Processing Units (GPGPUs)). sPEGG aims to markedly accelerate the analysis of eco-evolutionary dynamics using individual-based, forward-time population genetics models levereging cost-effective GPGPUs. sPEGG allows investigators to simulate the evolution of multi-locus traits under a wide array of ecological (e.g., age or spatially structured populations, multi-species communities) and genetic (e.g., mutation, variable linkage maps) scenarios. Below, we describe the overall design and key features of sPEGG. We then illustrate its performance gains relative to equivalent CPU implementations for three case studies. 

\newpage
\begin{figure}[h!]
   \centering
   \includegraphics[width=0.6\linewidth] {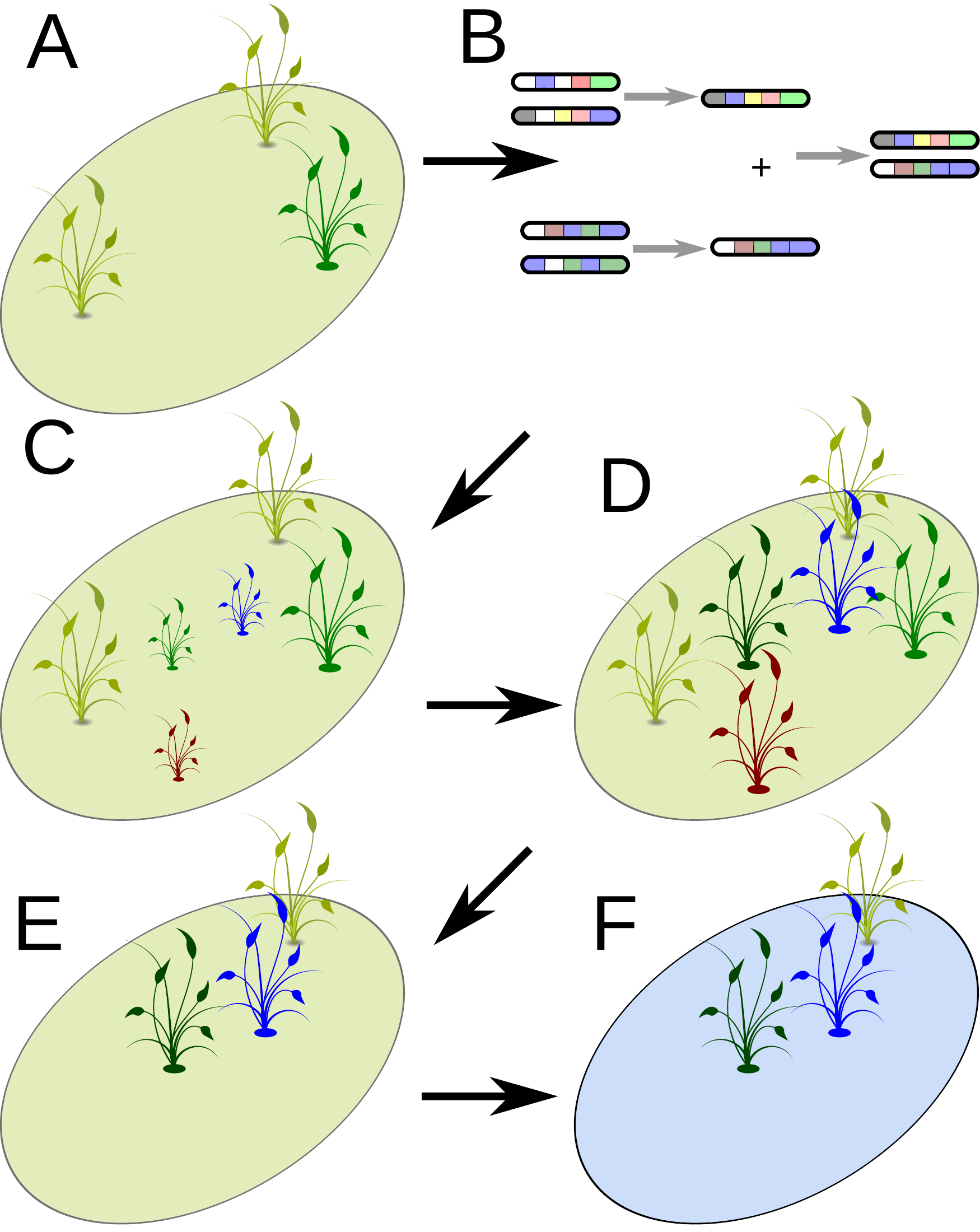}
\caption*{Figure 1. Calculations performed in sPEGG during each time step. Parents are identified (A), and genetic information is transmitted to offspring by subroutines implementing recombination and mutation (B). User-supplied functions map the resulting phenotypic distribution to newborns (C). A selective environment (here, substrate color in each panel) updates individual phenotypes (including mortality risk) (D) and survivors are identified (E). The relevant environmental variables variables are updated to reflect the phenotype distribution (F), thereby closing the eco-evolutionary feedback loop. Though not illustrated, gene flow resulting from migration among patches can also be modelled in sPEGG.}
\end{figure}

In sPEGG, the following calculations are performed at each step on each individual in each of the $n$ species. First, the eligible parents are identified according to rules supplied by the user, and the number of neonates are determined (Fig 1A). Second, parents are assigned to neonates, and genetic information is transmitted from the assigned parents to their offspring (Fig 1B). Third, based on the genetic information, the offspring's phenotype at birth is determined, based on a user-supplied genotype-phenotype map (Fig 1C). We highlight that sPEGG allows users to specify any genotype-phenotype map. Thus, in addition to an additive effect among loci, non-additive genetic effects such as dominance and epistasis, as well as maternal effects, environmental effects and gene $\times$ environment effects which characterize the trait of interest (\citealt{hendry13}) can be readily accommodated. Fourth, the individual's phenotypes (including their probability of survivorship) are updated according to the selective environment (which includes ecological interactions) that the user specifies (Fig 1D). Fifth, the survivors are identified and dead individuals are removed from the model (Fig 1E). A further step may involve updating the prevailing environmental conditions according to the phenotypic distribution of survivors, thereby completing the eco-evolutionary feedback loop (Fig 1F). The ability to simulate migration and gene flow in spatially-structured environments is also supported. sPEGG could be viewed as analogous to a numerical ordinary differential equation solver. As in those solvers, sPEGG still requires the user to provide some code describing the species-specific calculations, but the core computations above are carried out by the software. The online Methods describe sPEGG's design in greater detail.

sPEGG accommodates alternative mating systems (random, shared preference across individuals, or assortative;  monogamy or polygynandry). The transfer of genetic information can mimic recombination and sexual reproduction in diploid species or asexual reproduction, and can involve mutation and arbitrary linkage patterns. Importantly, all of these calculations, and more, are done across individuals using parallel algorithms, as described in the online Methods.

To illustrate the versatility and performance benefits of sPEGG, we use it to build IBMs to investigate three classic problems in evolutionary ecology. We first consider a very simple model of a randomly mating, diploid species with overlapping generations, in which the per-capita density-independent birth and death rate are quantitative traits whose numerical values are determined additively by five independently segregating loci subject to mutation in their allelic values (case study 1). Here, the population simply evolves to minimize density-independent mortality and maximize the density-independent birth rate (Supplementary Figure S1). Figure 2A shows that, as more individuals are simulated the total simulation speed increases approximately 10-fold compared to an equivalent simulation running entirely on a single CPU core.

\begin{figure}[h!]
   \centering
   \includegraphics[width=\linewidth] {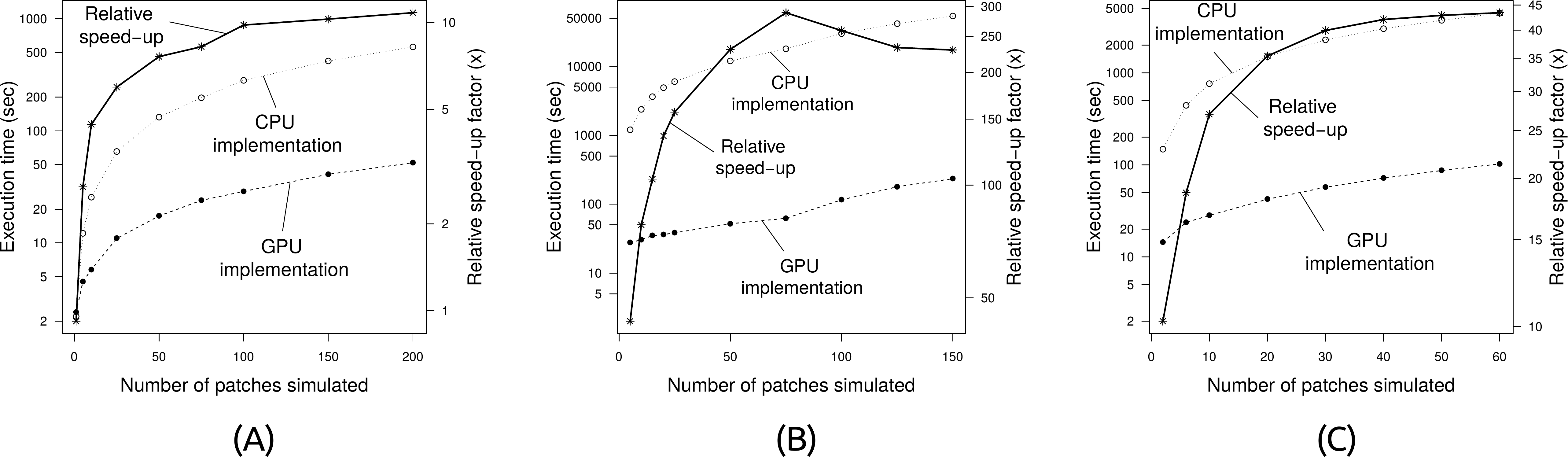}
\caption*{Figure 2. Performance comparisons as functions of the problem size (number of patches simulated) in terms of the number of seconds it takes to execute the simulations for the CPU version of sPEGG (dotted lines, open circles) and the GPU version (dashed lines, filled circles), as well as the speed up of sPEGG relative to a serial implementation (solid lines, asterisks). Performance comparisons are based on 250 iterations of steps represented in Figure 1 in the main text for both implementations. (A) Results from case study 1, which models evolution in fecundity and density-independent survivorship. (B) Results from case study 2, which models the evolution of neonate size in a physiologically structured model for a species that experiences an ontogenetic niche shift. (C) Results from case study 3, which models coevolving consumer-resource metacommunities linked by migration. In all panels, each patch contains approximately 50,000 individuals of each species; thus, a simulation of 100 patches simulates approximately $5\times10^6$ individuals in (A), (B) and approximately $10^7$ individuals in (C).}
\end{figure}

Next, we use sPEGG to model a size- and physiologically-structured (e.g., \citealt{persson98}), sexually reproducing species in which individuals of different sizes complete for two dynamic, biological resources (case study 2). Individuals transition from utilizing one resource type to utilizing another as they increase in size. We model the evolution of offspring size at birth, a trait which we assume (i) is inversely related to clutch size, and (ii) is controlled additively by a finite number of loci. Offspring size is then modeled to evolve to maximize individual fitness, subject to the individual-level constraints of size-specific survivorship, resource-use, and fecundity (Supplementary Information S2). Figure 2B illustrates how when the evolutionary question is addressed by a complex IBM, sPEGG can accelerate the simulation on a GPGPU by a factor of over 200x.

As a final case study, we model a spatially-structured, consumer-resource system in which the resource species and the consumer species coevolve (case study 3). We also model anti-predator defense to be costly for the resource. We apply sPEGG to analyze coevolution in such a system, investigating how a spatial gradient in the cost of anti-predator defense drives coevolutionary dynamics (Supplementary Information S3). Figure 2c shows that the performance improvements provided by sPEGG are higher for case study 3 than for case study 1 (up to approximately a factor of 44x), and not as high as the performance improvements for case study 2. 

The GPGPU versions of all case studies show accelerations of up to 1-2 orders of magnitude compared to the serial implementations. The differences in performance between the case studies reflect differences in their biological complexity. The fraction of the code's execution time spent performing calculations evaluating fitness, as opposed to reading and writing data from memory (albeit within the confines of GPU memory), is higher for case studies 2 and 3 than for case study 1 (where the speed of the simulation is limited instead by the speed of memory access). We note, however, that even in complicated IBMs, memory access may eventually prove increasingly taxing as more patches are simulated (Fig. 2B). Our comparative analysis of the three case studies thus identifies the biological circumstances under which sPEGG is likely to be significantly more efficient than serial implementations.   

Our results show that the performance-improvements provided by sPEGG can best be exploited when calculating how ecological processes affect individual fitness, delivering performance improvements comparable to a small cluster for our most complicated case-study model. In addition to being the first general framework for constructing and accelerating evolutionary individual-based models on GPGPUs, sPEGG also represents the first resource using a generalized, forward time population genetics model to study multi-species evolution in ecological communities. Thus, sPEGG may be an especially attractive framework for using system-specific, mechanistic models to elucidate the interplay between ecological and evolutionary dynamics. 

\section*{Acknowledgments}
This research was funded in part by a Chair's Fellowship from the Department of Ecology and Evolutionary Biology at the University of California, Los Angeles to K.W.O and a Complex Systems Scholar Award from the James S. McDonnell Foundation and  NSF grant  DEB 1457815 to P.A. We would like to thank G. Lin for discussion and assistance with some of the programming. 

\bibliographystyle{cbe}	
\bibliography{myrefs11}

\begin{thebibliography}{}

\bibitem[\protect\astroncite{Carvajal-Rodr{\'{\i}}guez}{2010}]{carvajal10}
{\sc Carvajal-Rodr{\'{\i}}guez, A.} 2010.
\newblock {Simulation of Genes and Genomes Forward in Time}.
\newblock {\em Current Genomics} 11:58--61.

\bibitem[\protect\astroncite{DeAngelis and Mooij}{2005}]{deangelis05}
{\sc DeAngelis, D. and Mooij, W.~M.} 2005.
\newblock {Individual-based modeling of ecological and evolutionary processes}.
\newblock {\em Annual Review of Ecology, Evolution and Systematics}
  36:147--168.

\bibitem[\protect\astroncite{Fuller and Millett}{2011}]{fuller11}
{\sc Fuller, S.~H. and Millett, L.~I.} 2011.
\newblock {Computing Performance: Game Over or Next Level?}
\newblock {\em Computer} 44:31--38.

\bibitem[\protect\astroncite{Futuyma}{2010}]{futuyma10}
{\sc Futuyma, D.~J.} 2010.
\newblock {Evolutionary constraint and ecological consequences}.
\newblock {\em Evolution} 64:1865--1884.

\bibitem[\protect\astroncite{Hartig et~al.}{2011}]{hartig11}
{\sc Hartig, F., Calabrese, J.~M., Reineking, B., Wiegand, T., and Huth, A.}
  2011.
\newblock {Statistical inference for stochastic simulation models--theory and
  application.}
\newblock {\em Ecology Letters} 14:816--827.

\bibitem[\protect\astroncite{Hendry}{2013}]{hendry13}
{\sc Hendry, A.~P.} 2013.
\newblock {Key questions in the genetics and genomics of eco-evolutionary
  dynamics}.
\newblock {\em Heredity} 111:456--466.

\bibitem[\protect\astroncite{Hendry}{ress}]{hendry16}
{\sc Hendry, A.~P.} In press.
\newblock {Eco-evolutionary dynamics}.
\newblock Princeton University Press, Princeton, NJ.

\bibitem[\protect\astroncite{Hoban et~al.}{2012}]{hoban12}
{\sc Hoban, S., Bertorelle, G., and Gaggiotti, O.~E.} 2012.
\newblock {Computer simulations: tools for population and evolutionary
  genetics}.
\newblock {\em Nature Reviews Genetics} 13:110--122.

\bibitem[\protect\astroncite{Pacala et~al.}{1996}]{pacala96}
{\sc Pacala, S.~W., Canham, C.~D., Saponara, J., {Silander Jr}, J.~A., Kobe,
  R.~K., and Ribbens, E.} 1996.
\newblock {Forest models defined by field measurements: estimation, error
  analysis and dynamics}.
\newblock {\em Ecological Monographs} 66:1--43.

\bibitem[\protect\astroncite{Persson et~al.}{1998}]{persson98}
{\sc Persson, L., Leonardsson, K., de~Roos, a.~M., Gyllenberg, M., and
  Christensen, B.} 1998.
\newblock {Ontogenetic scaling of foraging rates and the dynamics of a
  size-structured consumer-resource model.}
\newblock {\em Theoretical population biology} 54:270--293.

\bibitem[\protect\astroncite{Post and Palkovacs}{2009}]{post09}
{\sc Post, D.~M. and Palkovacs, E.~P.} 2009.
\newblock {Eco-evolutionary feedbacks in community and ecosystem ecology:
  interactions between the ecological theatre and the evolutionary play}.
\newblock {\em Philosophical Transactions of the Royal Society B: Biological
  Sciences} 364:1629--1640.

\bibitem[\protect\astroncite{Schoener}{2011}]{schoener11}
{\sc Schoener, T.~W.} 2011.
\newblock {The Newest Synthesis: Understanding the Interplay of Evolutionary
  and Ecological Dynamics}.
\newblock {\em Science} 331:426--429.

\bibitem[\protect\astroncite{Stapley et~al.}{2010}]{stapley10}
{\sc Stapley, J., Reger, J., Feulner, P. G.~D., Smadja, C., Galindo, J.,
  Ekblom, R., Bennison, C., Ball, A.~D., Beckerman, A.~P., and Slate, J.} 2010.
\newblock {Adaptation genomics: the next generation.}
\newblock {\em Trends in Ecology and Evolution} 25:705--712.

\bibitem[\protect\astroncite{Yoshida et~al.}{2003}]{yoshida03}
{\sc Yoshida, T., Jones, L.~E., Ellner, S.~P., Fussmann, G.~F., and Hairston,
  N.~G.} 2003.
\newblock {Rapid evolution drives ecological dynamics in a predator-prey
  system.}
\newblock {\em Nature} 424:303--6.

\end{thebibliography}


\begin{thebibliography}{}

\bibitem[\protect\astroncite{Booch}{1982}]{booch82}
{\sc Booch, G.} 1982.
\newblock {Object-oriented design}.
\newblock {\em ACM SIGAda Ada Letters} 1:64--76.

\bibitem[\protect\astroncite{Galassi et~al.}{2007}]{gsl}
{\sc Galassi, M., Davies, J., Theiler, J., Gough, B., Jungman, G., Alken, P.,
  Booth, M., and Rossi, F.} 2007.
\newblock {GNU Scientific Library Reference Manual, 3rd Ed.}
\newblock Network Theory.

\bibitem[\protect\astroncite{Harish and Narayanan}{2007}]{harish07}
{\sc Harish, P. and Narayanan, P.~J.} 2007.
\newblock {Accelerating large graph algorithms on the GPU using CUDA}.
\newblock {\em High Performance Computing–HiPC 2007} 4873:197--208.

\bibitem[\protect\astroncite{Hoberock and Bell}{2010}]{thrust}
{\sc Hoberock, J. and Bell, N.} 2010.
\newblock {Thrust: A Parallel Template Library}.

\bibitem[\protect\astroncite{Hruska}{2012}]{hruska12}
{\sc Hruska, J.} 2012.
\newblock {The death of CPU scaling: From one core to many — and why we’re
  still stuck}.
\newblock {\em ExtremeTech} .

\bibitem[\protect\astroncite{Martin}{2002}]{martin02}
{\sc Martin, R.~C.} 2002.
\newblock {Agile Software Development, Principles, Patterns, and Practices}.
\newblock Alan Apt Series. Prentice Hall.

\bibitem[\protect\astroncite{Nickolls et~al.}{2008}]{nickolls08}
{\sc Nickolls, J., Buck, I., Garland, M., and Skadron, K.} 2008.
\newblock {Scalable parallel programming with CUDA}.
\newblock {\em Queue} 6:40--53.

\bibitem[\protect\astroncite{Nyland et~al.}{2007}]{nyland07}
{\sc Nyland, L., Harris, M., and Prins, J.} 2007.
\newblock {Fast n-body simulation with {\{}CUDA{\}}}.
\newblock {\em GPU gems} 3:677--695.

\bibitem[\protect\astroncite{Owens et~al.}{2008}]{owens08}
{\sc Owens, J.~D., Houston, M., Luebke, D., Green, S., Stone, J.~E., and
  Phillips, J.~C.} 2008.
\newblock {GPU Computing}.
\newblock {\em Proceedings of the IEEE} 96:879--899.

\bibitem[\protect\astroncite{Pelletier et~al.}{2009}]{pelletier09}
{\sc Pelletier, F., Garant, D., and Hendry, A.} 2009.
\newblock {Eco-evolutionary dynamics}.
\newblock {\em Philosophical Transactions of the Royal Society B: Biological
  Sciences} 364:1483--1489.

\bibitem[\protect\astroncite{Rumbaugh et~al.}{1990}]{rumbaugh90}
{\sc Rumbaugh, J.~R., Blaha, M.~R., Lorensen, W., Eddy, F., and Premerlani, W.}
  1990.
\newblock {Object-oriented modeling and design}.
\newblock Prentice-Hall.

\bibitem[\protect\astroncite{Stallman}{2007}]{gpl3}
{\sc Stallman, R.} 2007.
\newblock {GNU General Public License v3}.

\bibitem[\protect\astroncite{Stepanov and Lee}{1994}]{stl}
{\sc Stepanov, A. and Lee, M.} 1994.
\newblock {The standard template library}.
\newblock Technical report, WG21/N0482, ISO Programming Language C++ Project.

\end{thebibliography}


\begin{thebibliography}{}

\bibitem[\protect\astroncite{Agresti}{2002}]{agresti02}
{\sc Agresti, A.} 2002.
\newblock Categorical Data Analysis.
\newblock Wiley-IEEE.

\bibitem[\protect\astroncite{Arino et~al.}{1998}]{arino98}
{\sc Arino, O., Hbid, M.~H., and {de la Parra}, R.~B.} 1998.
\newblock A mathematical model of growth of population of fish in the larval
  stage: Density-dependence effects.
\newblock {\em Mathematical Biosciences} 150:1--20.

\bibitem[\protect\astroncite{Bellinato and Bogliani}{1995}]{bellinato95}
{\sc Bellinato, F. and Bogliani, G.} 1995.
\newblock Colonial breeding imposes increased predation: experimental studies
  with herons.
\newblock {\em Ethology, Ecology and Evolution} 7:347--353.

\bibitem[\protect\astroncite{Blanckenhorn}{2005}]{blanckenhorn05}
{\sc Blanckenhorn, W.~U.} 2005.
\newblock Behavioral causes and consequences of sexual size dimorphism.
\newblock {\em Ethology} 111:977--1016.

\bibitem[\protect\astroncite{Broekhuizen et~al.}{1994}]{broekhuizen94}
{\sc Broekhuizen, N., Gurney, W. S.~C., Jones, A., and Bryant, A.~D.} 1994.
\newblock Modeling compensatory growth.
\newblock {\em Functional Ecology} 8:770--782.

\bibitem[\protect\astroncite{Brown et~al.}{2004}]{brown04}
{\sc Brown, J.~H., Gillooly, J.~F., Allen, A.~P., Savage, V.~M., and West,
  G.~B.} 2004.
\newblock {Toward a metabolic theory of ecology}.
\newblock {\em Ecology} 85:1771--1789.

\bibitem[\protect\astroncite{Bulmer}{1985}]{bulmer85b}
{\sc Bulmer, M.~G.} 1985.
\newblock The Mathematical Theory of Quantitative Genetics.
\newblock Clarendon Press, Oxford.

\bibitem[\protect\astroncite{B\"{u}rger}{2000}]{burger00}
{\sc B\"{u}rger, R.} 2000.
\newblock The Mathematical Theory of Selection, Recombination and Mutation.
\newblock Wiley Series in Mathematical and Computational Biology.

\bibitem[\protect\astroncite{Caley et~al.}{1996}]{caley96}
{\sc Caley, M.~J., Carr, M.~H., Hixon, M.~A., Hughes, T.~P., Jones, G.~P., and
  Menge, B.~A.} 1996.
\newblock Recruitment and the local dynamics of open marine populations.
\newblock {\em Annual Review of Ecology and Systematics} 27:477--500.

\bibitem[\protect\astroncite{Claessen et~al.}{2000}]{claessen00}
{\sc Claessen, D., de~Roos, A., and Persson, L.} 2000.
\newblock Dwarfs and giants: Cannibalism and competition in size-structured
  populations.
\newblock {\em American Naturalist} 155:219--237.

\bibitem[\protect\astroncite{Claessen and Dieckmann}{2002}]{claessen02}
{\sc Claessen, D. and Dieckmann, U.} 2002.
\newblock Ontogenetic niche shifts and evolutionary branching in
  size-structured populations.
\newblock {\em Evolutionary Ecology Research} 4:189--217.

\bibitem[\protect\astroncite{Costantino et~al.}{1997}]{costantino97}
{\sc Costantino, R.~F., Desharnais, R.~A., Cushing, J.~M., and Dennis, B.}
  1997.
\newblock Chaotic dynamics in an insect population.
\newblock {\em Science} 275:389--391.

\bibitem[\protect\astroncite{de~Roos and Persson}{2001}]{deroos01}
{\sc de~Roos, A.~M. and Persson, L.} 2001.
\newblock {Physiologically structured models - from versatile technique to
  ecological theory}.
\newblock {\em Oikos} 94:51--71.

\bibitem[\protect\astroncite{de~Roos and Persson}{2003}]{deRoos031}
{\sc de~Roos, A.~M. and Persson, L.} 2003.
\newblock {Competition in size-structured populations: mechanisms inducing
  cohort formation and population cycles}.
\newblock {\em Theoretical Population Biology} 63:1--16.

\bibitem[\protect\astroncite{Ellner and Rees}{2006}]{ellner06}
{\sc Ellner, S.~P. and Rees, M.} 2006.
\newblock Integral projection models for species with complex demography.
\newblock {\em American Naturalist} 167:410--428.

\bibitem[\protect\astroncite{Ernande et~al.}{2004}]{ernande04}
{\sc Ernande, B., Dieckmann, U., and Heino, M.} 2004.
\newblock Adaptive changes in harvested populations: plasticity and evolution
  of age and size at maturation.
\newblock {\em Proceedings of the Royal Society of London Series B-Biological
  Sciences} 271:415--423.

\bibitem[\protect\astroncite{Forrester}{1995}]{forrester95}
{\sc Forrester, G.~E.} 1995.
\newblock Strong density-dependent survival and recruitment regulate the
  abundance of a coral reef fish.
\newblock {\em Oecologia} 103:275--282.

\bibitem[\protect\astroncite{Hesse et~al.}{2008}]{hesse08}
{\sc Hesse, E., Rees, M., and M\"{u}ller-Sch\"{a}rer, H.} 2008.
\newblock Life-history variation in contrasting habitats: flowering decisions
  in a clonal perennial herb (\textit{{Veratrum} album}).
\newblock {\em American Naturalist} 172:E196--E213.

\bibitem[\protect\astroncite{Houle}{1996}]{houle96}
{\sc Houle, D.} 1996.
\newblock Comparing mutational variabilities.
\newblock {\em Genetics} 145:1467--1483.

\bibitem[\protect\astroncite{Lack}{1966}]{lack66}
{\sc Lack, D.~L.} 1966.
\newblock {Population studies in Birds}.
\newblock Oxford University Press.

\bibitem[\protect\astroncite{Leggett and {DeBlois}}{1994}]{leggett94}
{\sc Leggett, W.~C. and {DeBlois}, E.} 1994.
\newblock Recruitment in marine fishes: is it regulated by starvation and
  predation in the egg and larval stages?
\newblock {\em Netherlands Journal of Sea Research} 32:119--134.

\bibitem[\protect\astroncite{Lynch and Walsh}{1997}]{lynch97}
{\sc Lynch, M. and Walsh, B.} 1997.
\newblock Genetics and Analysis of Quantitative Traits.
\newblock Sinaur and Associates.

\bibitem[\protect\astroncite{Martin}{1993}]{martin93}
{\sc Martin, T.~E.} 1993.
\newblock Nest predation and nest sites.
\newblock {\em Bioscience} 43:523--532.

\bibitem[\protect\astroncite{Morris and Doak}{2002}]{morris02}
{\sc Morris, W.~F. and Doak, D.~F.} 2002.
\newblock Quantitative Conservation Biology: Theory and Practice of Population
  Viability Analysis.
\newblock Sinauer, {Sunderland, MA}.

\bibitem[\protect\astroncite{Moses et~al.}{2008}]{moses08}
{\sc Moses, M.~E., Hou, C., Woodruff, W.~H., West, G.~B., Nekola, J.~C., Zuo,
  W.~Y., and Brown, J.~H.} 2008.
\newblock Revisiting a model of ontogenetic growth: {Estimating} model
  parameters from theory and data.
\newblock {\em American Naturalist} 171:632--645.

\bibitem[\protect\astroncite{Mousseau and Roff}{1997}]{mousseau87}
{\sc Mousseau, T.~A. and Roff, D.~A.} 1997.
\newblock Natural selection and the heritability of fitness components.
\newblock {\em Heredity} 59:181--197.

\bibitem[\protect\astroncite{Mueller et~al.}{1991}]{mueller91}
{\sc Mueller, L.~D., Guo, P.~Z., and Ayala, F.~J.} 1991.
\newblock {Density-dependent natural selection and trade-offs in life history
  traits.}
\newblock {\em Science} 253:433--435.

\bibitem[\protect\astroncite{Parker and Begon}{1986}]{parker86}
{\sc Parker, G.~A. and Begon, M.} 1986.
\newblock Optimal egg size and clutch size: Effects of environment and maternal
  phenotype.
\newblock {\em The American Naturalist} 128:573--592.

\bibitem[\protect\astroncite{Persson et~al.}{1998}]{persson98}
{\sc Persson, L., Leonardsson, K., de~Roos, a.~M., Gyllenberg, M., and
  Christensen, B.} 1998.
\newblock {Ontogenetic scaling of foraging rates and the dynamics of a
  size-structured consumer-resource model.}
\newblock {\em Theoretical population biology} 54:270--293.

\bibitem[\protect\astroncite{Price and Schluter}{1991}]{price91}
{\sc Price, T. and Schluter, D.} 1991.
\newblock On the low heritability of life-history traits.
\newblock {\em Evolution} 45:853--861.

\bibitem[\protect\astroncite{Roff}{2001}]{roff01}
{\sc Roff, D.~A.} 2001.
\newblock {Life History Evolution}.
\newblock Sinauer Associates.

\bibitem[\protect\astroncite{Roff and Emerson}{2006}]{roff06}
{\sc Roff, D.~A. and Emerson, K.} 2006.
\newblock Epistasis and dominance: evidence for differential effects in
  life-history versus morphological traits.
\newblock {\em Evolution} 60:1981--1990.

\bibitem[\protect\astroncite{Sale}{1976}]{sale76}
{\sc Sale, P.~F.} 1976.
\newblock The effect of territorial adult pomacentrid fishes on the recruitment
  and survival of juveniles on patches of coral rubble.
\newblock {\em Journal of Experimental Marine Biology and Ecology} 24:297--306.

\bibitem[\protect\astroncite{Shepherd and Cushing}{1980}]{shepherd80}
{\sc Shepherd, J.~G. and Cushing, D.~H.} 1980.
\newblock A mechanism for density-dependent survival of larval fish as the
  basis of a stock-recruitment relationship.
\newblock {\em ICES Journal of Marine Science} 39:160--167.

\bibitem[\protect\astroncite{Stearns}{1976}]{stearns76}
{\sc Stearns, S.~C.} 1976.
\newblock Life history tactics: a review of the ideas.
\newblock {\em Quarterly Review of Biology} 51:3--47.

\bibitem[\protect\astroncite{Trivers}{1972}]{trivers72}
{\sc Trivers, R.~L.} 1972.
\newblock Parental investment and sexual selection.
\newblock {\em In} B. Campbell (ed.), Sexual selection and the descent of man,
  pp. 136 -- 178. Aldine.

\bibitem[\protect\astroncite{{Van Doorn} and Dieckmann}{2006}]{van06}
{\sc {Van Doorn}, G.~S. and Dieckmann, U.} 2006.
\newblock {The long-term evolution of multilocus traits under
  frequency-dependent disruptive selection}.
\newblock {\em Evolution} 60:2226--2238.

\bibitem[\protect\astroncite{Werner and Gilliam}{1984}]{werner84}
{\sc Werner, E.~E. and Gilliam, J.~F.} 1984.
\newblock {The ontogenetic niche and species interactions in size structured
  populations}.
\newblock {\em Annual Review of Ecology and Systematics} 15:393--425.

\bibitem[\protect\astroncite{West et~al.}{2001}]{west01}
{\sc West, G.~B., Brown, J.~H., and Enquist, B.~J.} 2001.
\newblock A general model for ontogenetic growth.
\newblock {\em Nature} 413:628--631.

\bibitem[\protect\astroncite{White and Warner}{2007}]{white07}
{\sc White, J.~W. and Warner, R.~R.} 2007.
\newblock Safety in numbers and the spatial scaling of density-dependent
  mortality in a coral reef fish.
\newblock {\em Ecology} 88:3044--3054.

\bibitem[\protect\astroncite{Wilson and Osenberg}{2002}]{wilson02}
{\sc Wilson, J. and Osenberg, C.~W.} 2002.
\newblock Experimental and observational patterns of density-depedent
  settlement and survival in the marine fish \textit{Gobiosoma}.
\newblock {\em Oecologia} 130:205--215.

\bibitem[\protect\astroncite{Winemiller and Rose}{1992}]{winemiller92}
{\sc Winemiller, K.~O. and Rose, K.~A.} 1992.
\newblock Patterns of life-history diversification in north-american fishes -
  implications for population regulation.
\newblock {\em Canadian Journal of Fisheries and Aquatic Sciences}
  49:2196--2218.

\bibitem[\protect\astroncite{Wootton}{1998}]{wootton98}
{\sc Wootton, R.~J.} 1998.
\newblock Ecology of teleost fishes.
\newblock Chapman and Hall.

\end{thebibliography}


\begin{thebibliography}{}

\bibitem[\protect\astroncite{Bazzaz et~al.}{1987}]{bazzaz87}
{\sc Bazzaz, F.~a., Chiariello, N.~R., Coley, P.~D., and Pitelka, L.~F.} 1987.
\newblock Allocating resources to reproduction and defense.
\newblock {\em BioScience} 37:58--67.

\bibitem[\protect\astroncite{Bohannan and Lenski}{1997}]{bohannan97}
{\sc Bohannan, B. J.~M. and Lenski, R.~E.} 1997.
\newblock Effect of resource enrichment on a chemostat community of bacteria
  and bacteriophage.
\newblock {\em Ecology} 78:2303--2315.

\bibitem[\protect\astroncite{Hassell}{1978}]{hassell78}
{\sc Hassell, M.~P.} 1978.
\newblock {The dynamics of arthropod predator-prey systems.}
\newblock {\em Monographs In Population Biology} pp. 1--237.

\bibitem[\protect\astroncite{Nuismer et~al.}{2007}]{nuismer07}
{\sc Nuismer, S.~L., Ridenhour, B.~J., and Oswald, B.~P.} 2007.
\newblock {Antagonistic coevolution mediated by phenotypic differences between
  quantitative traits}.
\newblock {\em Evolution} 61:1823--1834.

\bibitem[\protect\astroncite{Rasmann and Agrawal}{2011}]{rasmann11a}
{\sc Rasmann, S. and Agrawal, A.~A.} 2011.
\newblock Evolution of specialization: a phylogenetic study of host range in
  the red milkweed beetle (\textit{Tetraopes tetraophthalmus}).
\newblock {\em The American Naturalist} 177:728--737.

\end{thebibliography}
\newpage 

\section*{Supplementary Material}
Supplementary Figure S1. The evolutionary trajectory of the traits modeled in case study 1.

\noindent Supplementary Information S2. Description of eco-evolutionary model used in case study 2.

\noindent Supplementary Information S3. Description of eco-evolutionary model used in case study 3.

\noindent Supplementary Figure S4. Description of the species class used in sPEGG models.

\noindent Supplementary Figure S5. Pseudo-code illustrating the difference between the serial  and parallel implementations of individual-based models in sPEGG. 

\end{document}


\begin{figure}[h!]
   \centering
   \begin{subfigure}[b]{0.49\textwidth}
    \centering
   \includegraphics[width=\linewidth] {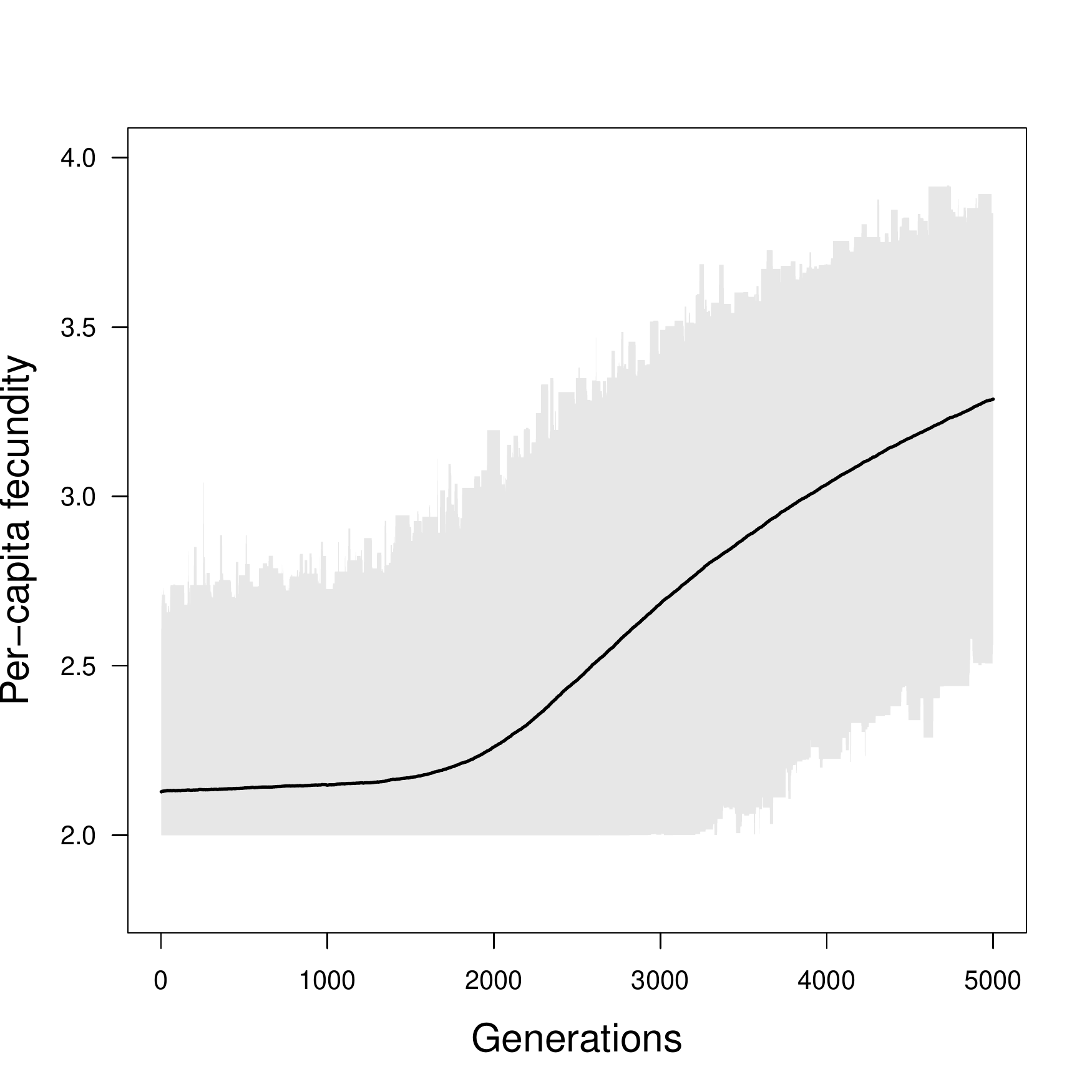}\quad
	\caption*{(A)}

	\end{subfigure}
   \begin{subfigure}[b]{0.49\textwidth}
    \centering
   \includegraphics[width=\linewidth] {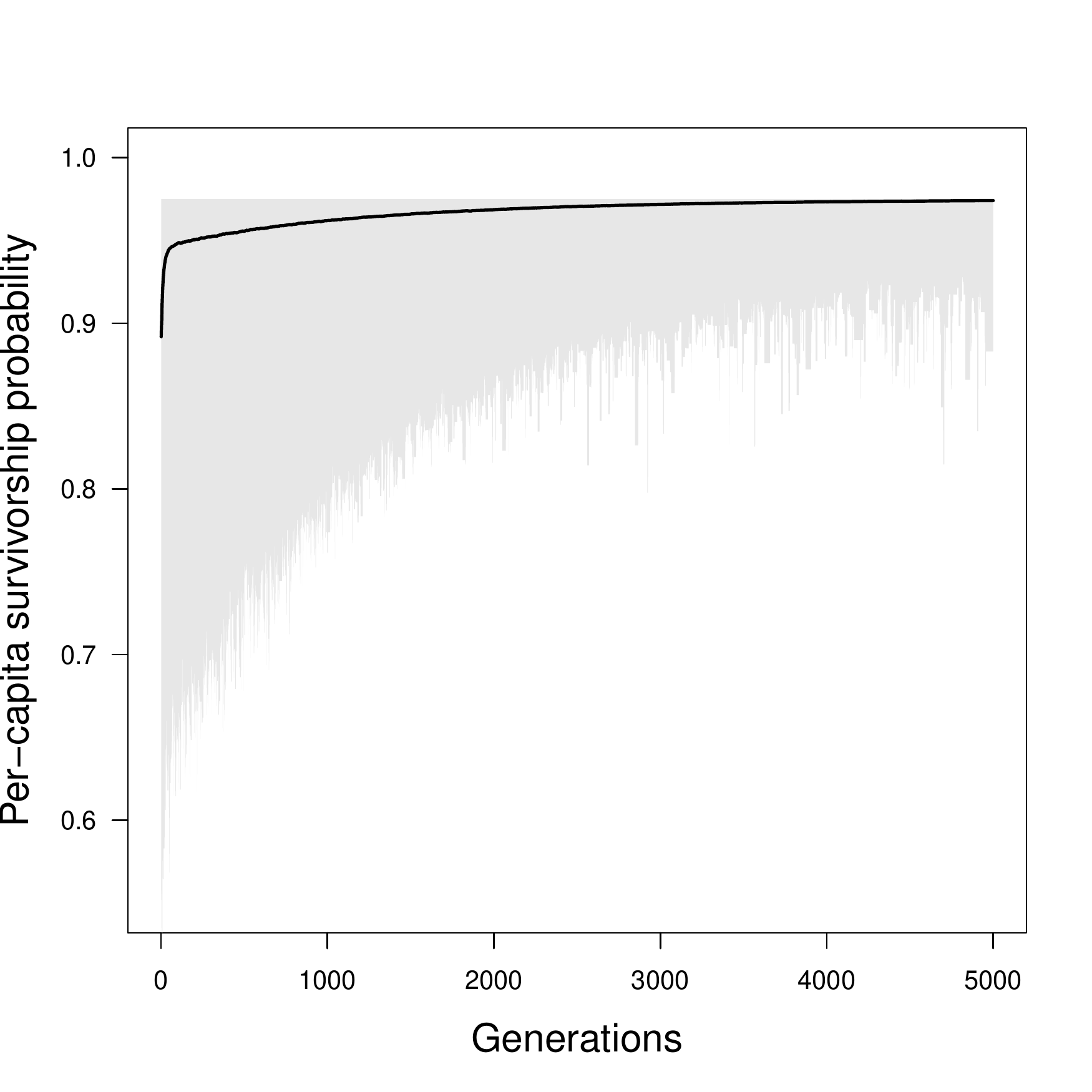}
	\caption*{(B)}
  	\end{subfigure}
\caption*{Figure S1. The evolutionary trajectory of the average traits determining per-capita fecundity (A) and the per-time step per-capita survivorship probability (B). To prevent individuals from living forever (i.e., to have a survivorship probability of one), the maximum evolvable per-time step survivorship has an asymptote at 0.975, which corresponds to an average lifespan of approximately 40 time steps. In each panel, the grey region represents the maximum and minimum trait values in each deme. For further details, see the main text.}

\end{figure}



  \renewcommand{\theequation}{S2-\arabic{equation}}
 
\setcounter{equation}{0}  
\section*{Supplementary Information S2 - Model details for simulation of evolving neonate mass in the face of ontogenetic niche shifts}

In many species, the ability of individuals to exploit different resources changes as individuals grow in size or develop across life stages, a process referred to as an ontogenetic niche shift (\citealt{werner84}). When consumption by conspecifics causes resource limitation, ontogenetic niche shifts can impose distinct patterns of density-dependence during an individual's life cycle. Such density-dependence could affect the ability of newborns to survive, grow and mature, potentially affecting patterns of life history evolution (e.g., \citealt{mueller91}). For instance, evolutionary changes in neonate mass are subject to a trade-off between selection for larger clutch sizes and increased offspring survivorship (e.g. \citealt{lack66}, \citealt{parker86} and \citealt{roff01}). However, the strength of this trade-off itself can depend upon prevailing patterns of resources available for juvenile growth (necessary for escaping vulnerable size classes) and adult reproduction (which partially governs the number of offspring a parent can birth or sire). These patterns of resource availability at different life stages may, in turn, depend on the strength of an ontogenetic niche shift individuals experience.

To understand the effect of ontogenetic niche shifts on the evolution of neonate mass, we use sPEGG to model a physiologically-structured (e.g., \citealt{persson98} and \citealt{deroos01}), sexually reproducing consumer population whose individuals utilize and compete over two dynamic, biological resources. We assume maximum resource consumption rates increase allometrically (\citealt{brown04}), but that individuals transition from utilizing one resource type to utilizing another as they increase in size. We model offspring size at birth as an evolvable trait controlled additively by a finite number of loci of large effect. We assume survivorship increases monotonically with individual size (e.g., \citealt{werner84}), but that parents with larger neonates also birth fewer offspring (e.g., \citealt{parker86}). 

\subsection*{Model overview}
We consider a size-structured, sexually reproducing consumer population. Our model has two key components - an ecological component and a genetic component. The ecological component describes the individual's interactions with their environment. We model how differential reproductive success and survivorship (fitness) among individuals emerge from these diverse interactions. The genetic component specifies the genetic distribution for neonate mass, and how the genetic distribution changes between generations. 

The model's dynamics are iterated on a discrete time step. At each time step, the model cycles through all individuals to determine their fates. We assume that the organisms reproduce once every $T$ time-steps; thus, each time-step can be thought of, for instance, as a day for a specie's with an annual breeding season. Our model applies to organisms with overlapping generations and seasonal reproduction such as insects, marine invertebrates, and vertebrates that live in seasonal environments. We also partition individual somatic mass $W$ into reversible ($Y$) and irreversible ($X$) mass to explicitly study how energetic constraints affect individual reproductive decisions. 

The model's parameters characterize the selective pressures and the genetic, energetic and ecological constraints on the population. Its ecological dynamics are driven by four key processes: resource consumption, somatic growth, mating, and mortality. Below, we describe how each process is modelled in further detail. Table S2-1 provides a summary of the model's ecological components. 

\subsection*{Resource Consumption}
We consider two dynamic, biological resources which the consumers utilize and for which they compete. Resource availability can regulate individual somatic growth, and ultimately affects individual survivorship and reproduction. Thus, considering two resources allows us to model situations where newborns and large adults potentially experience different regulation regimes, and hence permit different forms of resource scarcity to operate during an individual's lifetime. 

The instantaneous resource-consumption rate $E_{i,{\tau},j}$ of individual $i$ of resource $j = 1,2$ at time $\tau$ is a function of its body size (somatic mass $W_{i,\tau}$) and resource density $R_{\tau,j}$:
\begin{linenomath}
\begin{equation}
E_{i,\tau} (R_{\tau,j}, K_j, W_{i,\tau}) =  h(R_{\tau,j},K_j) \pi(W_{i,\tau}, j) \alpha W_{i,\tau}^{\gamma} 
\end{equation}
\end{linenomath}
where $\pi(W_{i,\tau}, j)$ describes the proportion of a individual $i$'s diet that consist of resource type $j$, and $h(R_{\tau,j}, K_j)$ describes the proportion by which an individual's resource consumption decreases when the resource density $R_{\tau,j}$ falls short of the maximum daily consumption rate (attained when the resource density is at carrying capacity $K_j$). 

We assume that if the resources are at their carrying capacities, then an individual's instantaneous resource consumption rate depends allometrically on its body size, where the parameters $\gamma$ and $\alpha$ are an allometric exponent and an allometric constant scaling consumption rates, respectively (e.g., \citealt{west01}). 

The size-independent function $h(R_{\tau,j},K_j)$ describes how much an individual's consumption of resource $j$ decreases as the becomes more scarce. For example, if resource $j$ has carrying capacity $K_j$ and consumer individuals display a Holling type-II functional response with attack rate $a$ and handling time $T_H$, then $h(R_{\tau,j},K_j)$ describes how resource limitation constrains consumption, i.e., $h(R_{\tau,j},K_j) =\frac{a R_{\tau,j}/K_j}{1 + a T_H R_{\tau,j}/K_j}$. 

We assume that the amount of a given resource of type $j$ a consumer utilizes depends on the consumer's size. Such ontogenetic niche shifts are common in nature (e.g., \citealt{werner84}). For example, many fish are planktivorous when small, but shift to piscivory as they increase in size (e.g., \citealt{claessen02}). The function $\pi(X_{i,\tau}, j)$ describes the proportion of individual $i$'s diet that consist of resource type $j$. The function $\pi(X_{i,\tau},j)$ is modeled to depend on both anatomical structures such as gape-size and body length that contribute to irreversible mass  (e.g., \citealt{werner84}). It is also determined by a constant $u$ that governs what fraction of the adult resource base very small individuals consume, and the steepness $p$ of the curve as individuals grow. Thus, $\pi(X_{i,\tau},j)$ is defined by the following logistic equations:

\begin{linenomath}
\begin{equation}
\label{eqn:pi}
\pi(W_{i, \tau}, j) = \left\{ \begin{array}{ll}
\frac{1}{1 + \exp(- p(l[W_{\tau,i}] - u l[W_{max}]))}, & \textrm{if } j = 1 \\
1 - \frac{1}{1 + \exp(- p(l[W_{\tau,i}] - u l[W_{max}]))}, & \textrm{if } j = 2 
\end{array} \right.
\end{equation}
\end{linenomath}

l[] is a function mapping irreversible mass to body length(e.g., \citealt{agresti02} and \citealt{claessen02}). We highlight that when $\pi(X_{i, \tau}, j)$ is invariant with $X_{i,\tau}$ (e.g., $p$ small), this describes a situation where resource scarcity affects all individuals similarly during their life-time.

Resource variability is driven by both extrinsic and intrinsic sources of variability. Extrinsic sources of variability, such as climactic fluctuations, affect the per-capita intrinsic growth rate $r_j^{\prime}$ of resource $j$. By contrast, temporal variation in resource utilization by the consumer generates variability in resource levels that is intrinsic to the system.

We assume that the dynamics of resource $j$ are described by intrinsic, density-dependent growth and predation by the consumer (e.g., \citealt{claessen00}) and are governed by the following Beverton-Holt-like growth curve:
\begin{linenomath}
\begin{equation}
R_{\tau+1,j} = \frac{r_j^{\prime} R_{\tau,j}}{1+(R_{\tau,j})/K_j}  - \sum_{i=1}^{N_{\tau}} E_{i,\tau,j} (R_{\tau,j})
\end{equation}
\end{linenomath}
where $E_{i,0,j} (R_{0,j})$ is the consumption rate of resource $j$ by the individual consumer $i$ at the start of the time step, and $N_{\tau}$ is the population size of consumers at the beginning of the time step. 

We model stochastic fluctuations in the resource dynamics due to extrinsic factors, such as climactic variability, by allowing $r_j^{\prime}$ to vary from time step to time step by drawing the actual value of $r_j^{\prime}$ used in the evaluation of $R_{t,j}^{\star}$ from a normal distribution with mean $r_j$ and standard deviation $e_{r,j}$. Our model does not consider externally imposed deterministic fluctuations. Hence, temporal variability in resource abundance and the consumer's population dynamics are emergent properties of the underlying ecological processes we model, as well as the stochastic nature of the model itself. 

\subsection*{Somatic Growth}
In our model, somatic growth generates size structure. We model two components of somatic growth: growth in irreversible mass $X$ and growth in reversible mass $Y$. An individual's irreversible, or structural, mass $X$ consists of compounds such as organ and skeletal tissue that cannot be starved away (e.g, \citealt{broekhuizen94} and \citealt{deroos01}). Irreversible mass can be viewed as a surrogate for body length, which does not decrease even under starvation conditions (e.g., \citealt{broekhuizen94}). By contrast, an individual's reversible mass is determined by energy reserves such as lipids and gonadal tissue in mature individuals that can be starved away. Reversible mass is partitioned into the mass of storage tissue $Y$, and, in mature individuals, the mass of gonadal tissue, $G$ (e.g., \citealt{broekhuizen94}, \citealt{persson98}, \citealt{deroos01}) (Fig. 1B). Hence,
\begin{linenomath}
\begin{equation}
W = X + Y + G.
\end{equation}
\end{linenomath}
In females, gonadal mass $G$ consists largely of reproductive tissue, while in males, gonadal mass $G$ is interpreted as the amount of reversible mass allocated to reproduction through the loss of somatic mass incurred by competing with other males for access to mature females as well as the production of reproductive tissue.

We assume that individuals are born with the maximum reversible mass to irreversible mass ratio. We define an individual's condition, $c$, as the ratio of storage tissue mass to total somatic mass, i.e., $c=\frac{Y}{W}$. 

An individual consumer's size at maturity and the fraction of resources it allocates towards reproduction affects how it allocates its energetic intake between irreversible, reversible, and gonadal mass. Here, we seek to link resource consumption to an individual's physiological state mechanistically. An individual's physiological state (e.g., body size, condition, etc...) determines its performance (in particular, survivorship and reproduction). Therefore, explicating how metabolic constraints affect somatic growth is key to describing the feedback between an individual's environment and its physiological state.

A consumer with mass $W_{i,\tau}$ at time $\tau$ grows according to the following growth equation (e.g., \citealt{west01})
\begin{linenomath}
\begin{equation}
\label{eqn:growth1}
W_{i,\tau+1} = H(R_{\tau}, W_{i,\tau}) \alpha W_{i,\tau}^{\gamma}  - \delta W_{i,\tau},
\end{equation}
\end{linenomath}
where $\delta$ describes the metabolic rate per unit mass of consumer tissue, $H(R_{\tau}, W_{i,\tau}) = \sum_{i=1,2} (h(R_{\tau,j}, K_j) \pi(W_{i,\tau}))$ describes how resource scarcity ($R_{\tau}$) at time $\tau$ reduces the growth rate of an individual of size $W_{i,\tau}$, and the parameters $\gamma$ and $\alpha$ are an allometric constant and an allometric exponent scaling consumption. We therefore assume that when the mass derived from consumption exceeds metabolic costs, any surplus mass is invested in somatic growth. For taxa with indeterminate growth, $W_{max}$ represents an asymptotic body size which individuals approach at a decelerating rate. For taxa with determinate growth, $W_{max}$ simlpy defines the body size at which the somatic growth rate is zero. We assume $\gamma \leq 1 $ so that there is an asymptotic maximum size $W_{max}$ beyond which matabolic costs render further somatic growth impossible (for empirical values of $\gamma$, see, e.g., \citealt{moses08}).

Consumer individuals are assumed to be born with the maximum ratio $q_J$ between reversible mass and irreversible mass. We note that at birth, an individual's reversible mass does not include gonadal mass - instead, reversible mass consists of storage tissue such as fats. Upon maturation, reversible mass includes the mass of gonadal tissue as well as storage tissue. Hence, the corresponding maximum ratio between reversible mass and irreversible mass for adults is given by $q_A > q_J$. Because reversible mass can be starved away, it represents a reserve that individuals can utilize during periods of resource limitation. Thus, when resources are not limiting, we assume consumer individuals will maintain the maximum ratio of reversible to irreversible mass. The fraction $\kappa$ of $ \Delta W_i = W_{i,\tau+1} - W_{i,\tau}$ that is allocated to irreversible mass depends on the ratio of reversible to irreversible mass; thus:
\begin{linenomath}
\begin{equation}
\label{eqn:growth1}
X_{i,\tau+1} = \kappa \frac{Y_{i,\tau}}{X_{i,\tau}} \Delta W_i + X_{i,\tau+1},
Y_{i,\tau+1} = (1-\kappa \frac{Y_{i,\tau}}{X_{i,\tau}}) \Delta W_i + Y_{i,\tau+1}.
\end{equation}
\end{linenomath}

Hence, $\kappa$ describes a constraint on how much an individual can allocate resource intake towards fending off starvation risk (by improving condition) instead of reducing other forms of density-independent mortality by increasing irreversible mass.

In mature individuals, a fixed proportion $\rho_i$ of $1-\kappa \frac{Y_{i,\tau}}{X_{i,\tau}}) \Delta W_i$ is set aside for reproduction. For example, in female animals this takes the form of allocation to reproductive tissue mass $G_i$, while in males this represents, for example, the mass loss associated with searching and securing female mates as well as the energy expended in producing gonadal tissue. For simplicity, we assume the mass of reproductive tissue in juveniles is negligible. Thus, at the end of each time step, reproductive tissue mass for individual $i$ is given by:
\begin{linenomath}
\begin{eqnarray}
\label{eqn:allocations}
G_{i,\tau+1} &=& I_f \times (Y_{i,\tau} - X_{i,\tau} q_A)
\end{eqnarray}
\end{linenomath}
where $I_f = 0$ if the individual is immature, $I_f = 1$ if the individual is mature. The growth equations above are based on the assumption that all reproductive tissue mass from the previous time step was spent on reproduction during that time step. Following reproduction, the gonadal mass $G_{i,\tau+1}$ is deducted from the individual's reversible mass $Y_{i,\tau}$.

Equations (\ref{eqn:allocations}) summarize how energetic constraints govern the feedback between an individual's environment and its physiological state. In particular, they describe how resource use affects an individual's reproductive success (which depends on gonadal mass $G$) and survivorship (which depends on irreversible mass $X$ and condition $c=Y/(X+Y+G)$).

\subsection*{Reproduction}
Individual consumers breed once every $T$ time steps, and we assume random mating between males and females, conditioned on gonadal mass. If, prior to mating, an individual consumer $i$'s irreversible mass $X_i$ is larger than its size at maturity ($\sigma_i$), then individual $i$ is considered to be mature. 

The number $F$ of fertilized eggs in the population is determined by three variables: (1) the average neonate mass of offspring of female $i$, $\Omega_i + q_J \Omega_i$, where $q_J$ is the maximum ratio of reversible mass to irreversible mass, (2) the reproductive tissue mass of female $i$ $G_i$, and (3) $N_{f,t}$, the total number of mature females in time step $t$. Then,
\begin{linenomath}
\begin{equation}
\label{eqn:fecundity}
F = \sum_{i=1}^{N_{f,t}} \frac{G_i}{\Omega_i + q_J \Omega_i}
\end{equation}
\end{linenomath}
The present modeling framework applies to both viviparous and oviparous organisms. However, for simplicity, we discuss model development in terms of an oviparous organism. 

For each egg, the mother and father are drawn at random from mature males and females in the population. The probability that the egg comes from female $i$ is a function of the female's relative gonadal mass $G_i$ in the population,
\begin{linenomath}
\begin{equation}
\label{eqn:Femalerepr}
\Pr(\textrm{female $i$ produces an egg} | G_i) = (\frac{\frac{G_i}{\Omega_i + q_J \Omega_i }}{F})^{v_f},
\end{equation}
\end{linenomath}
where $v_f$ describes the severity fo reproductive skew among females.

\indent Similarly, the probability that a mature male $i$ fertilizes a given egg is a function of its irreversible mass, $X_i$, relative to the mass of other mature males in the population. In particular, our model reflects the common observation that body size (often measured as length, which depends on irreversible mass) is positively correlated with reproductive success in males (e.g., \citealt{trivers72}, \citealt{blanckenhorn05}), and a male's relative structural mass is incorporated into calculating the probability of fertilization as: 
\begin{linenomath}
\begin{equation}
\Pr(\textrm{male $i$ fertilizes an egg} |X_i, G_i) = \frac{X_i^{v_m}}{\sum_{j=1}^{N_{m,t}}(X_j)^{v_m}}
\end{equation}
\end{linenomath}
where $N_{m,t}$ is the total number of mature males, and $v_m$ determines how strongly a male's reproductive value increases with its relative irreversible mass. The parameter $v_m$ can describe a host of biological processes, including female preference for larger males and the superior fighting ability of larger males. High values of $v_m$ characterize populations where reproductive success is strongly correlated with male body size, while lower values of $v_m$ characterize populations where male reproductive success is similar across a range of body sizes. 

\subsection*{Mortality}
\indent At any given time step, the number $\Phi$ of newborns that survive to the juvenile stage depends on the population's total egg production (e.g., \citealt{wootton98}). Several mechanisms can contribute to such density-dependent mortality at the earliest life stages (\citealt{shepherd80}). For instance, in many organisms early life stages (e.g., larvae) are particularly vulnerable to predators that can be attracted to large aggregations of eggs or hatchlings (\citealt{martin93}, \citealt{bellinato95}, and \citealt{white07}). Furthermore, when hatchlings or early larvae form dense aggregations over small spatial scales, such aggregations can lead to the rapid exhaustion of locally available resources (e.g., \citealt{leggett94}, \citealt{arino98}). Finally, in many organisms,  dispersal occurs at very early stages of the life cycle. In such situations, the ability of newborns to become established and survive to maturation at suitable sites depends, in part, on the fraction of sites already occupied by breeding adults, which in turn affects the value of $F$ (eq. (\ref{eqn:fecundity}) e.g., \citealt{caley96}). For example, in several coral reef fishes, the density of adults is inversely correlated with the survivorship of juveniles, possibly because of competition faced by juveniles for suitable shelter sites that protect individuals from predation or the availability of desirable feeding sites (e.g., \citealt{sale76}, \citealt{forrester95}, \citealt{caley96}, and \citealt{wilson02}). When these mechanisms interact, as when limited resource availability stunts somatic growth and thereby keeps new borns from growing out of vulnerable size classes, the relationship between the number of surviving newborns and total fecundity is described by classical stock-recruitment curves (\citealt{shepherd80}).

Here we model the density-dependent recruitment $\Phi$ according to the function:
\begin{linenomath}
\begin{equation}
\Phi = \frac{\nu}{1+N_f},
\end{equation}
\end{linenomath}
where $\nu$ is a constant and $N_f$ is the total number of reproductive females. For viviparous organisms, $\Phi$ can be interpreted as the fraction of offspring that survive past a weaning period. 

\indent In addition to density-dependent mortality of newborns, individual survivorship from one time step to the next is a function of the individual's irreversible mass. We assume that the probability that an individual survives from one time step to the next is described as
\begin{linenomath}
\begin{eqnarray}
\label{eqn:survivorship1}
\Pr(\textrm{individual $i$ survives} | X_{i,t}) = \frac{1}{1+\exp(-\beta_2 (X_{i,t} - \beta_1))}
\end{eqnarray}
\end{linenomath}
This functional form results from using standard logistic regression to model survivorship probability as a function of size, as is frequently recommended (e.g., \citealt{morris02}, \citealt{ellner06}, and \citealt{hesse08}). The parameter $\beta_1$ represents the mass (scaled by the maximum irreversible mass $W_{max}$) at which the mortality risk is equal to 1/2. The parameter $\beta_2$ characterizes the steepness of the survivorship curve around this point. This functional form is quite flexible, and depending on the values of $\beta_1$ and $\beta_2$ can describe an exponential increase in survival with increased irreversible mass, a sigmoidal increase in survival with larger irreversible mass, a relatively linear increase in mortality risk with irreversible mass, or a rate of mortality that is largely independent of body size.

We further assume that individuals also suffer an additional source of mortality due to starvation. We assume that as individual condition worsens, starvation risk increases and survival decreases at an exponential rate. The survivorship function thus depends on individual condition and somatic mass as:
\begin{linenomath}
\begin{eqnarray}
\label{eqn:survivorship}
  \Pr(\textrm{individual $i$ survives at time $t$} | X_{i,t},c_{i,t}) = (1-\exp(-\beta_s Y_{i,t}/X_{i,t})) \times \frac{1}{1+\exp(-\beta_2 (X_{i,t} - \beta_1))}. 
\end{eqnarray}
\end{linenomath}
The parameter $\beta_s$ characterizes how survivorship increases exponentially with improving condition. Because resource availability governs somatic growth rates and determines an individual's body size and condition, eq. (17) accounts for the potential for density-dependent processes to affect survivorship throughout an individual's life cycle.

Size-specific mortality links an individual's physiological state to its performance. Size-specific mortality can directly select for different neonate sizes, as well as generate population fluctuations and induce variability in resource availability (e.g.,  \citealt{stearns76}, \citealt{costantino97}, \citealt{deRoos031} and \citealt{ernande04}). In turn, these patterns of temporal variability could subsequently affect the evolution of life history syndromes (e.g., \citealt{winemiller92}).

\subsection*{Genetics}
In our model, the fitness of individuals emerges from their interactions with other individuals and their environment. The evolutionary response of the population to selection on individuals depends on the genetic distribution among individuals. Moreover, the genetic distribution, in turn, is determined by the combined effects of mutation, recombination, and selection. In individual-based models, describing the population's genetic distribution requires we specify the distribution of genotypes among individuals. Thus, specifying how multiple life history traits evolve requires explicit consideration of the genetics underlying the life history traits. Below, we describe how we model the genetics and transmission of the life history traits.

We analyze evolution in the mean irreversible mass $\Omega$ of offspring at birth (neonate mass). While survivorship often increases monotonically with individual size (e.g., \citealt{werner84}), parents that give birth to larger neonates give birth to fewer offspring (e.g., \citealt{parker86}). Thus, focusing on neonate mass allows us to study a trait that directly evolves in response to a trade-off between clutch size and offspring survivorship (e.g. \citealt{lack66} and \citealt{roff01}). 

We assume that offspring size at birth $\Omega$ is a quantitative traits whose genetic values are additively determined by $N_\Omega$ trait-specific loci. The genome is diploid, individuals reproduce sexually, and there is free recombination between all loci. Pleiotropy is absent from the model, and the allelic values of the loci vary continuously over the biologically feasible range of the life history trait. To describe the evolution of this life history trait, we track the dynamics of the alleles and loci underlying the trait explicitly. Such an explicit, multi-locus framework allows us to characterize and account for changes in the genetic distribution of the trait over time. All loci were treated as autosomal and freely recombining with other loci. The allelic values of the loci could change by mutations and the effect of mutations on the allelic values are assumed to be Gaussian distributed. We note that when the number of trait-specific loci approaches infinity and individual allelic effects decline to zero, these assumptions allow our model to recover the classical infinitesimal model used in quantitative genetics (e.g., \citealt{bulmer85b}). 

Prior to reproduction, each parent produces haploid gametes consisting of half the parent's alleles (e.g., \citealt{van06}). Mutations occur with a probability $\mu$ at each locus. If a mutation occurs at a locus, the new allelic value is drawn from a normal distribution with the mean at the allelic value prior to mutation and a standard deviation given by $\varpi \times$ the mean initial allelic value. The gametes from both parents fuse to produce an offspring's diploid genome. The offspring's genetic value at each locus is given by the midpoint of the parental gamete's allelic values at that locus. Thus, the offspring's genetic value for $\Omega$ is the sum of the genetic values across all $N_\Omega$ loci. 

Once an individual's genotypic value is additively determined, its phenotypic value is determined first by drawing a random variable, $z_J$ from a normal distribution with the mean given by the genotypic value and a trait-specific standard deviation $\varrho_J$. In effect, determining the phenotypic value in this way is analogous to specifying the residual variance for the trait (i.e., the difference between the trait's phenotypic and additive genetic variances - e.g., \citealt{houle96}). Thus, if $A_{i,f}, A_{i,m}$ denote the allelic value at locus $i$ an individual inherited from its female and male parent, respectively, and $N(M,\varrho)$ describes a normal distribution with mean $M$ and standard deviation $\varrho$, the individual's phenotypic value $z_J$ is a random variable given as:
\begin{linenomath}
\begin{eqnarray}
\label{eqn:phenval}
z_J \sim N(\displaystyle\sum_{i=1}^{N_J} \frac{1}{2} (A_{i,f} + A_{i,m}), \varrho_J).
\end{eqnarray}
\end{linenomath}

Life history traits appear to generally have low heritabilities (e.g., \citealt{mousseau87} and \citealt{price91}). Nevertheless, there is some question about the relative importance of dominant and epistatic effects among loci for life history traits (e.g., \citealt{roff06}) or whether environmental, gene-by-environment interactions, and developmental processes account for the low heritability in life history traits (e.g, \citealt{price91}, \citealt{houle96}, \citealt{burger00}). Because $\varrho_J$ is shared across individuals, our model formulation effectively assumes most of the residual variance reflects the effects of developmental and (unspecified) environmental effects rather than nonadditive genetic components. 

Finally, the values of $z_J$ were transformed using standard approaches to ensure the expressed phenotypic values were biologically realistic. In particular, we set  
\begin{linenomath}
\begin{eqnarray}
\label{eqn:survivorship}
z_J^{\prime} = \exp(z_J),
\end{eqnarray}
\end{linenomath}
for neonate mass at birth. Thus, when the genetic values of these traits follow a normal distribution, the genetic component of their phenotypic values follow a lognormal distribution (e.g., \citealt{lynch97}).

\subsection*{Analysis and Results}
We vary the density-independent, per-capita recruitment rate of the resource consumed by smaller individuals to assess how shifting the relative availability of juvenile and adult resources affects the evolution of offspring size at birth.  The parameter values used in the simulations of this model employed in the main text are available at \nolinkurl{https://github.com/kewok/spegg}. We find that depending on the relative availability of these resources, the consumer population can evolve towards two potential evolutionary endpoints. The first endpoint occurs where individuals give birth to a large number of small offspring. The second endpoint occurs where individuals give birth to a small number of large offspring (Supplementary Fig. S2). If the juvenile resource is relatively scarce, then this selects against parents that give birth to small offspring, who can remain in vulnerable size-categories until they acquire sufficient resources for growth. By contrast, if the juvenile resource is more abundant, this permits relatively rapid juvenile growth even by small offspring. This relaxes the offspring survivorship-clutch size trade-off, favoring the evolution of smaller body sizes at birth. The model illustrates how differences in a phenotype of the resource species (in this case, per-capita density-independent recruitment) can cascade through ontogenetic changes in the consumer to select for distinct consumer life-history strategies.


\bibliographystyle{cbe}	
\bibliography{myrefs}

\newpage
\begin{singlespace}
\begin{table}
\begin{center}
\begin{tabular}{l lp{3cm}}
\hline 
\multicolumn{2}{c} {Definition of State Variables and Individual-level Functions} \\ 
\hline
 $W_{i,t} \doteq$ Somatic mass of consumer individual $i$, which is the sum of irreversible (e.g., skeletal & \\
 \hspace{1cm} and muscle tissue) mass, $X_{i,t}$, and reversible mass (storage tissue, $Y_{i,t}$, and & \\
\hspace{1cm} mass of reproductive tissue (for mature individuals), $G_{i,t}$) at time $t$ & \\
 $F_t \doteq$ The number of fertilized eggs in time step $t$ &\\
 $N_{j,t} \doteq$ The number of mature individuals of sex $j$ in time-step $t$ &\\
 $\Omega_i \doteq$ The average irreversible mass of consumer individual $i$'s offspring &\\
 $\Phi_t \doteq$ The number of offspring surviving past the reproductive season in time-step $t$ &\\
 $R_{\tau,j} \doteq$ The biomass of resource $j$ at time $\tau$ &\\
 $N_{t} \doteq$ The number of consumers at time-step $t$&\\
 $E_{i,\tau} (R_{\tau,j}) \doteq$ The amount of resource $j$ consumed by consumer $i$ at time $\tau$ &\\
 $h(R_{\tau,j},K_j) \doteq $ Analogous to the functional response; describes the proportion by which consumer & \\ \hspace{1cm}individual $i$'s consumption of resource type $j$ is reduced by resource scracity. & \\
 $H(R_{\tau}, W_{i,\tau}) \doteq $ The effect of resource scarcity on consumer growth. \\
\hline
\multicolumn{2}{c} {Model Specification: Individual Level Processes}  \\ \hline
 Maturity condition for consumer $i$: $I_f = 1\textrm{ if }X_{i,t} \geq \sigma_i$, 0 otherwise & \\
 $\Pr(\textrm{mature female $i$ is mother of an egg} | G_i) = \frac{\frac{G_{i,t}}{\Omega_i + q_J \Omega_i }}{F_t}$ & \\
 $\Pr(\textrm{mature male \textit{i} fertilizes an egg} |X_i, G_i) = \frac{(X_{i,t})^{v_m}}{\sum_{j=1}^{N_{m,t}}(X_{j,t} )^{v_m}}$  & \\
 $\Pr(\textrm{individual \textit{i} survives between \textit{t}, \textit{t}+1} | X_{i,t},Y_{i,t}) = (1-\exp(-\beta_s Y_{i,t}/X_{i,t})) \times \frac{1}{1+\exp(-\beta_2 (X_{i,t} - \beta_1))}$  & \\
 $\frac{E_{i,\tau}(R_{\tau,j})}{\sum_{k=1}^{2} E_{i,\tau}(R_{\tau,k})} =\pi(W_{i, \tau}, j) = \left\{ \begin{array}{ll}
\frac{1}{1 + \exp(- p (W_{\tau,i} - u W_{max}))}, & \textrm{if } j = 1 \\
1 - \frac{1}{1 + \exp(- p (W_{\tau,i} - u W_{max}))}, & \textrm{if } j = 2 
\end{array} \right. $ \\

 $E_{i,\tau} (R_{\tau,j}) =  h(R_{\tau,j},K_j) \pi(W_{i,\tau}, j) \alpha W_{i,\tau}^{\gamma}$  & \\ 
$W_{i,t+1} =  H(R_{\tau}, W_{i,\tau}) \alpha W_{i,\tau}^{\gamma}  - \delta W_{i,\tau}$\\
$\Delta W_i = W_{i,t+1} - W_{i,t}$\\
$X_{i,\tau+1} = \kappa \frac{Y_{i,\tau}}{X_{i,\tau}} \Delta W_i + X_{i,\tau+1}$ & \\
$Y_{i,\tau+1} = (1-\kappa \frac{Y_{i,\tau}}{X_{i,\tau}}) \Delta W_i + Y_{i,\tau+1}$ \\
$G_{i,t+1} = I_f \times (Y_{i,\tau} - X_{i,\tau} q_A)$ &  \\
\hline
\multicolumn{2}{c} {Model Specification: Population Level Processes} \\ \hline
  $F_t = \sum_{i=1}^{N_{f,t}} \frac{G_{i,t}}{\Omega_i + q_J \Omega_i }$ & \\
  $\Phi_t = \frac{\kappa_1 F_t}{1 + (\frac{F_t}{\kappa_2}^{\kappa_{3}})}$ & \\
 $N_{t} = \Phi_t +N_{t-1} -$ Individuals that  die between ($t-1$) and $t$ \\ 
\hline
\end{tabular}
\caption*{Table S-2: Definition of the state variables in the ecological component of the model and a summary of the model's dynamics. The text provides justifications of the functional forms and a detailed formulation of the model. See text for the definitions of the parameters.}
\end{center}
\end{table}
\end{singlespace}

\newpage
\begin{figure}[h!]
   \includegraphics[width=\linewidth] {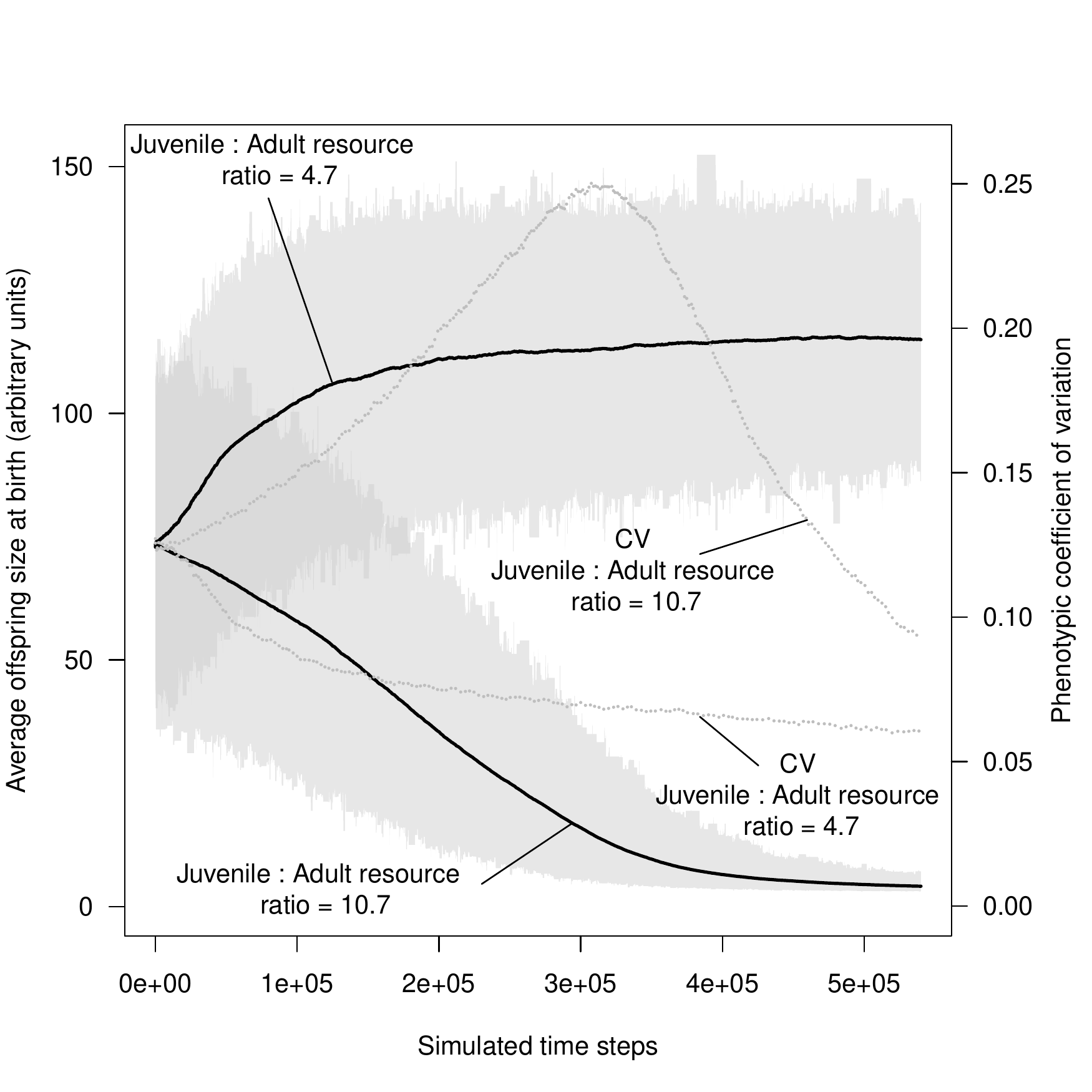}
\end{figure}
Supplementary Figure S2. The effect of relative resource scarcity of the juvenile resource on the evolution of body size of an individual at birth. Depending on how scarce juvenile resources are, alternative evolutionary trajectories in average body size at birth (black lines) and the phenotypic variance (illustrated here with the coefficient of variation) for neonate size (grey lines) are possible. The grey region represents the maximum and minimum trait values in each deme. For further details, see the main text.



\renewcommand{\theequation}{S3-\arabic{equation}}
 
\setcounter{equation}{0}  
\section*{Supplementary Information S3 - Model details for simulating a co-evolving resource-consumer metacommunity}
We model a $P$ patch meta-community where each patch is linked by migration to its neighboring patch. Within each patch, individuals of  the consumer species $C$ interact with individuals of the resource species $R$. We model coevolutionary dynamics in systems where consumers with a given phenotypic value are most effective at exploiting resources with a particular resource trait value, but may be poor at exploiting resources with other trait values (i.e., phentoypic matching - e.g., \citealt{nuismer07}). For instance, an herbivorous insect may evolve specific enzymes permitting it to exploit a novel plant resource, while the plant species may evolve alternative chemical defenses against the herbivore (\citealt{rasmann11a}). Alternatively, the ability of a resource to defend itself, and the ability of a consumer to exploit the resource, may itself be a composite trait of many distinct phenotypes (e.g., behavioral as well as morphological defenses), with consumer individuals varying in their ability to exploit different facets of resource defense (e.g., some consumers may be particularly good at identifying resource hiding sites, while others may have greater ability to puncture resource defenses such as shells).

In our model, the potential individual fecundity $w_{F,C}(t)$ of an individual consumer varies over time. An individual's breeding value determines the genetic component of this trait. $w_{F,C}(t)$ is also affected by how successfully an individual consumer is able to exploit the resource species. A consumer's ability to successfully exploit the resource, in turn, depends on two quantities: first, the individual's genetically determined exploitation ability, $\alpha$, and second, the phenotypic distribution of a resource defense trait $\delta$.  This adds a further source of genetic variation in individual fecundity, as well as an environmental effect that varies according to the prevailing distribution of the resource's phenotypes. At a given time $t$, the consumer's fecundity can be given by:

\begin{align}
w_{F,C}(t) &= \xi k \int_{y \in \{\delta\}} \frac{\exp(-(\alpha-y)^2)}{1 + \exp(-(\alpha-y)^2)} dy, \label{eqn:consumer_fecundity}
\end{align}
where $\xi$ is the breeding value characterizing the consumer's baseline fecundity and $k$ scales the effect of the interaction on consumers (it could thus represent, e.g., something like a conversion efficiency). Figure S3-A illustrates the per-capita effect of encounters between a given consumer and resource individual based on their phenotypic values. We note that removing the squared expression inside the exponential would describe a consumer-resource interaction with monotonic arms-race dynamics. 

The consumer's per-capita mortality rate $w_{M,C}$ is assumed to be constant during an individual consumer's's life; thus, we do not model reduction in survivorship through a failure to acquire resources (e.g., via starvation).  

Encounters with consumers are assumed to increase the mortality risk of the resource; we describe changes in the resource's mortality rate $w_{M,R}$ to depend on the extent to which the resource's trait is dissimilar to the distribution of consumer exploitation traits:

\begin{align}
w_{M,R}(t) &=  \gamma v \int_{x \in \{\alpha\}} \frac{\exp(-(x-\delta)^2)}{1 + \exp(-(x - \delta)^2)} dx,\label{eqn:resource_mortality}
\end{align}
where $\gamma$ is the breeding value characterizing the resource's baseline survivorship and $v$ scales the effect of the interactions on resource mortality.

We also model implicit exploitative competition among resource individuals. The outcome of such resource-resource competition determines the distribution of resource fecundities. We assume that the the resource's ability to defend itself against consumers comes at a cost to reproduction. Such a trade-off could arise, for instance, if energetic reserves need to be allocated to producing secondary compounds instead of fruit (e.g., \citealt{bazzaz87}; also \citealt{bohannan97}). Changes to resource fecundity are then modeled as:

\begin{align}
w_{F,R}(t) &= w_{F,R}(t) n_i \exp(-\rho \alpha^2),
\label{eqn:resource_fecundity}
\end{align}
where $n_i$ describes the aggregate effect of any encounters a resource individual $i$ may have with other resource individuals, and $\rho$ describes the severity of the trade-off between resource defense and resource fecundity. This entails that greater investment by the resource in any given anti-consumer strategy can entail greater fecundity costs.

Eqns. (\ref{eqn:consumer_fecundity} - \ref{eqn:resource_fecundity}) are mathematical abstractions in which a given individual of the focal species is assumed to encounter all possible individuals in the other species. In nature, however, encounters between individuals often result from Poisson processes (e.g., \citealt{hassell78}). Therefore, to capture the effects of contingent interspecific interactions on trait evolution, for a given individual of either species we instead simulate a random subset of possible encounters out of all possible species. The specific life history traits therefore change according to the cumulative interactions an individual experiences during a given time step, rather than summing over all possible interactions.

For the simulations of this model reported in the main text, we assumed that the phenotype governing consumer exploitation ability is controlled by three loci which we assume have additive effects on the consumer trait of interest. Similarly, the resource's defensive trait was assumed to be additively controlled by five loci. For both the consumer and resource species, genetic variations in the life history traits at birth were assumed to be minimal.

Figure S3-A illustrate how the strength $\rho$ of the tradeoff between resource defense and resource fecundity can govern the eco-evolutionary outcome of consumer-resource coevolution. When the tradeoff is weak, the resource can readily evolve anti-consumer defenses and drive the consumer extinct; when the tradeoff is strong, the resource can ill-afford to invest too heavily in a particular strategy and thus the manner in which it avoids predation shifts, enabling coevolutionary cycling. The consumer can only persist when the tradeoff faced by the resource is weak when there is sufficient migration from patches in which the tradeoff is strong. Figure S3-B show how the extent of migration between patches governs consumer persistence. Figure S3-B shows how when the tradeoff is weak, the consumer population struggles to persist. However, when cross patch dispersal is relatively high, the inflow of maladapted consumers prevents successful local adaptation and consumer persistence is somewhat more limited. The parameter values used in the simulations of this model employed in the main text are available at \nolinkurl{https://github.com/kewok/spegg}.


\bibliographystyle{cbe}	
\bibliography{myrefs}	
\newpage

\begin{figure}[h!]
   \centering
   \begin{subfigure}[b]{0.49\textwidth}
    \centering
   \includegraphics[width=\linewidth] {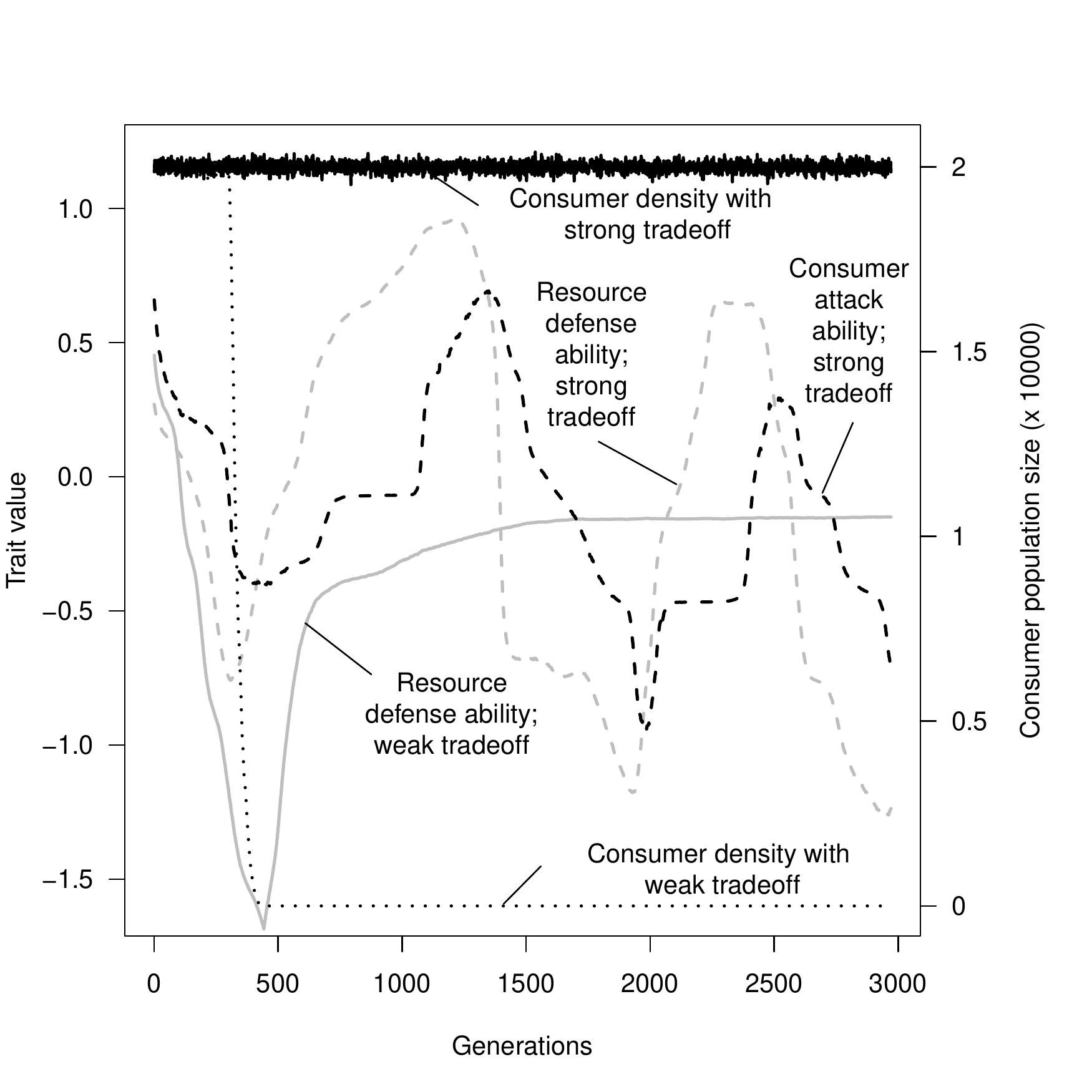}\quad
	\caption*{(A)}        
	\vspace{-1\baselineskip}

	\end{subfigure}
   \begin{subfigure}[b]{0.49\textwidth}
    \centering
   \includegraphics[width=\linewidth] {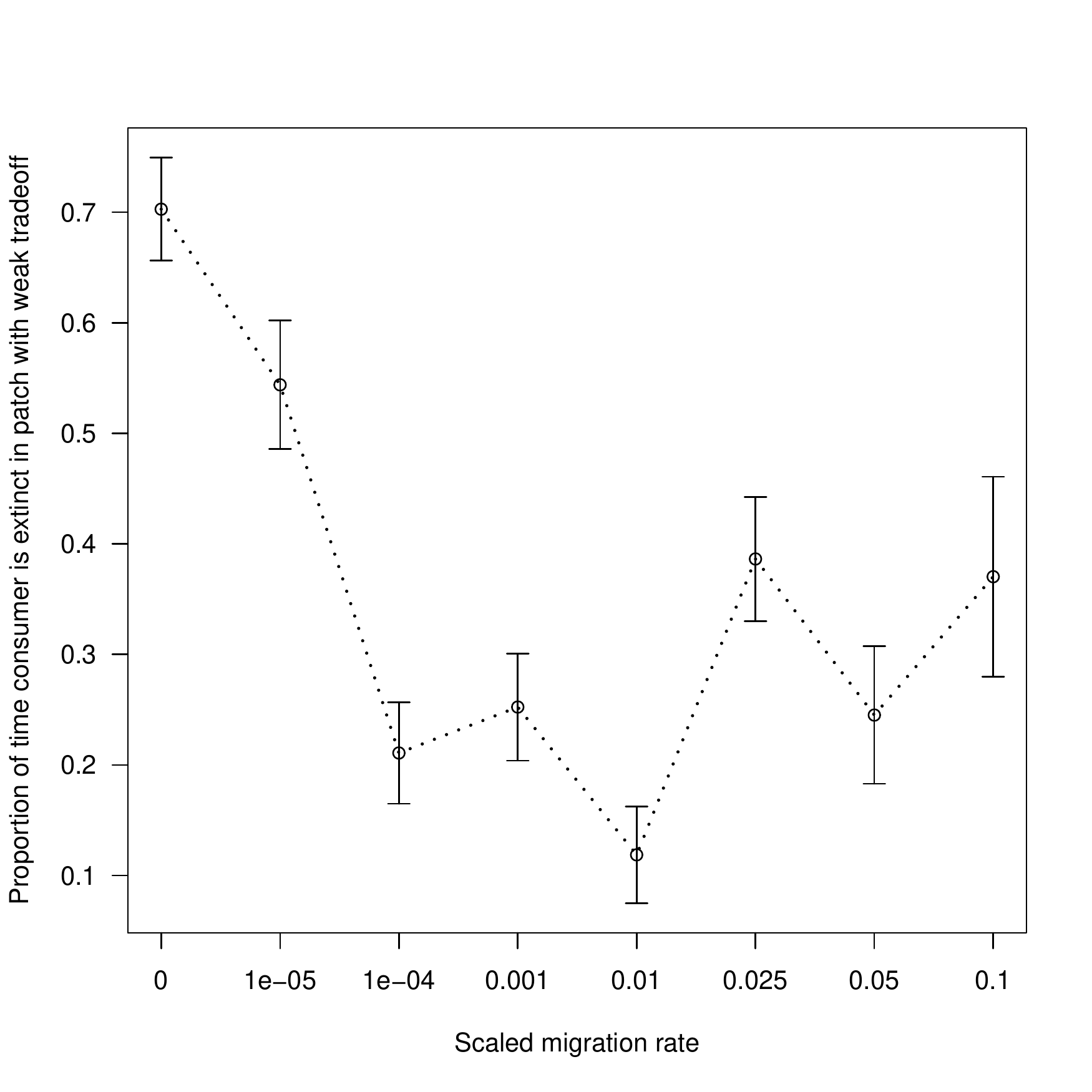}\quad
	\caption*{(B)}
        \vspace{-1\baselineskip}
  	\end{subfigure}
\caption*{}

\end{figure}
Supplementary Figure S3. (A) Coevolutionary dynamics in a consumer-resource community without migration. The traits affecting inter-specific interactions for the resource are subject to a pleiotropic constraint $\rho$ on fitness, which is modeled as reducing their per-capita fecundity following interactions with intra-specifics (i.e., their competitive ability) depending on how much the resource defense ability trait deviates from zero. The consumer, by contrast, is assumed to face no such constraint. Consumer-resource coevolution is therefore possible only in environments where the tradeoff faced by the resource is strong (black dashed line). (B) The effect of varying the scaled migration rate on consumer persistence, i.e., the proportion of time in which the consumer population is extinct, in the patch with relatively weak constraints on the resource defense trait. A biased island dispersal model is used, in which a proportion $\pi$ of migrants from each patch ultimately return to their patch of origin. Thus, the values on the abscissa in panel (B) represent the effective migration rate scaled by the propensity of individuals to attempt to leave their home patch. Error bars reflect the standard deviation across 15 replicate simulations.


\begin{figure}[h!]
\begin{center}
   \includegraphics[width=0.9\linewidth] {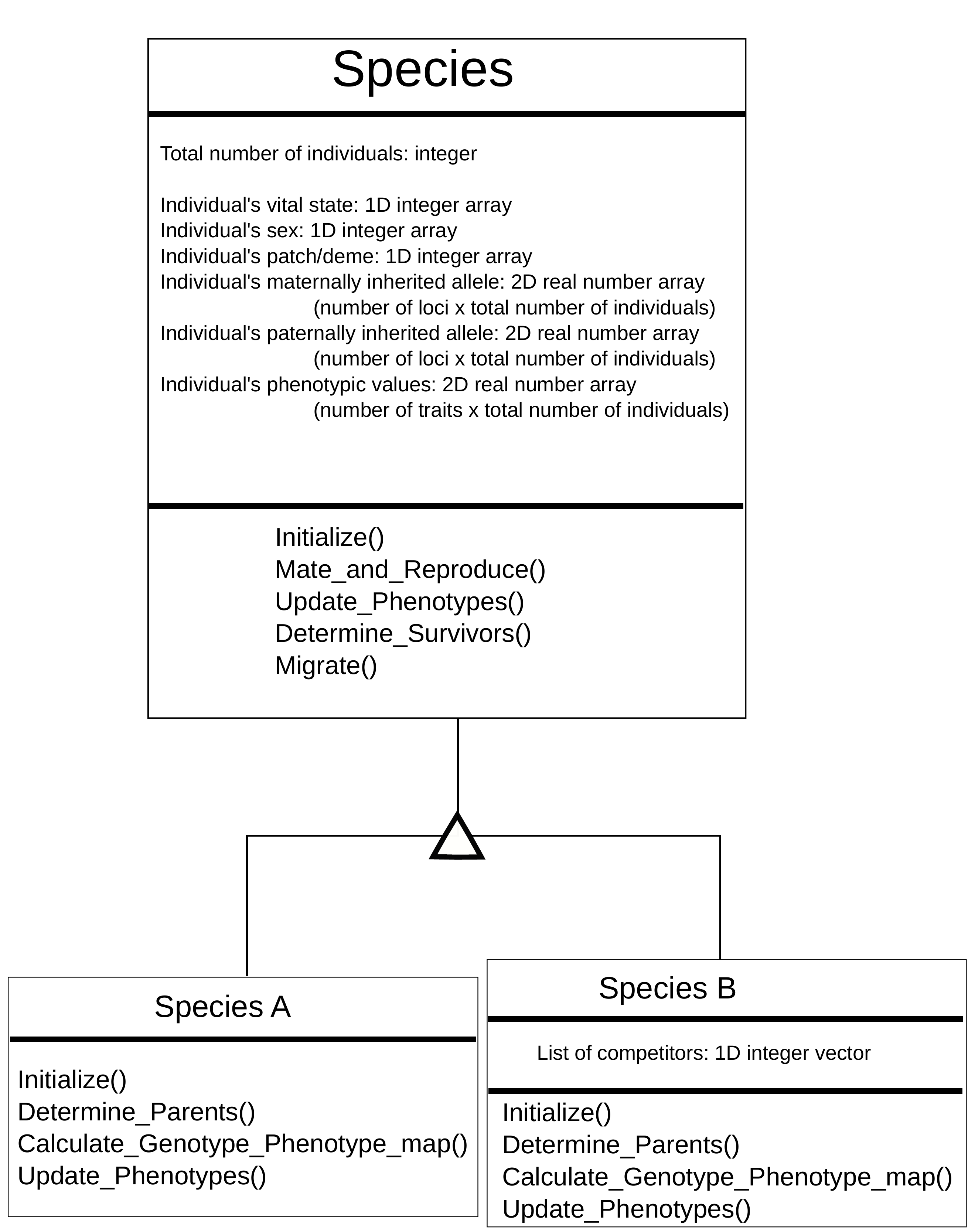}
\end{center}
\caption*{Figure S4.  A description of the base class for species that is used in sPEGG, and example relationships to derived, model-specific classes (species A and species B) using the object modelling technique. The first segment of each box speciefies the class name, the middle (if present) segment specifies the data members of the class, and the last segment lists the methods (functions) of the class. In this example, the methods \texttt{Initialize()}, \texttt{Determine\_Parents()}, \texttt{Calculate\_Genotype\_Phenotype\_map()} and \texttt{Update\_Phenotypes()} implement alternative, species-specific customizable routines that can potentially differ between species A and species B (e.g., different genetic architectures), and species B also contains a data member consisting of the indices of potential competitors (e.g., individuals in species A that potentially pre-empt resources from species B). The remaining attributes are common to species A and species B, as are the routines for simulating mating and reproduction, mortality, and migration.}
\end{figure}


\begin{algorithmic}
\Large{}
\For{$1 \leq S\leq $Number of Replicates}
\For{$1 \leq i\leq $Number of Individuals} 
	\State $parameter \gets $ Parameter values for 
	\State \hspace{1.75in} simulation $S$
	\State trait value of individual $i \gets foo(parameter)$
\EndFor
\EndFor
\end{algorithmic}
\begin{center}
\Large{(A)}
\end{center}

\begin{algorithmic}
\Large{}
\For{$1 \leq i\leq $ Number of individuals} 
 in\\
\hspace{1.5in} parallel
\Large{}
	\State $D \gets $ Replicate (patch) of individual $i$
	\State $parameter \gets $ Parameter values from
	\State \hspace{1.75in} deme $D$
	\State trait value of individual $i \gets foo(parameter)$
\EndFor
\end{algorithmic}
\begin{center}
\Large{(B)}\\
\end{center}

\begin{singlespace}
\noindent Supplementary Figure S5. Pseudo-code illustrating the primary difference between individual-based models following (A) a serial implementation and (B) a parallelized implementation using sPEGG. Each segment of pseudo-code represents calculations performed in Figure 1D of the main text to update the phenotypes of individuals using the user-supplied function \texttt{foo()} for each time step. In (A), when there are $n$ individuals and $p$ patches, then $np$ operations must be performed sequentially (one after another). By contrast, in (B), if there are $k$ parallel cores, then $k$ operations can be performed simultaneously, and under ideal conditions, evaluating the loop takes $np/k$ steps. In effect, the operations in (B) allow us to ``flatten'' the nested loop described in (A), enabling fine-grained parallelism of the individual-based model.
\end{singlespace}



\section*{Online Methods}
\subsection*{Application overview}
The core parallelization routines in sPEGG are implemented using Thrust (\citealt{thrust}), a highly optimized parallel algorithms library loosely resembling the widely used C++ standard template library (STL) (\citealt{stl}). To exploit GPGPU hardware, the present generation of Thrust uses CUDA (\citealt{nickolls08}). Thus, the GPGPU version of sPEGG can currently only run on CUDA-enabled NVIDIA hardware, although this may change as Thrust expands to support other GPGPU programming languages (e.g., OpenCL). Thrust also allows us to use multi-core CPU processors for parallellization. Currently, most consumer desktops have only a small number (generally 2-8) of CPU cores (e.g., \citealt{hruska12}). However, if CPU core counts increase dramatically at affordable prices while GPGPU hardware improvements stagnate, sPEGG's use of the Thrust library enables it to be well-positioned to exploit such hardware developments. 

In general, sPEGG relies on an object-oriented design (e.g., \citealt{booch82}, \citealt{rumbaugh90} and \citealt{martin02}). For instance, sPEGG models each species as a separate object, consisting of thrust vectors stored on GPGPU memory that describe the phenotypes and genotypes of individuals. The calculations are then performed in parallel on individual-level data stored in the species objects. Models created with sPEGG can then be written in standard C++. Supplementary Figure S4 illustrates how the species objects of sPEGG could be used in an IBM. 

sPEGG is released under the GNU Public License v3 (\citealt{gpl3}). The most recent version of sPEGG, further documentation, source code for the case studies (including their parameter values and initial conditions) and a short tutorial are available at \nolinkurl{https://github.com/kewok/spegg}.

\subsection*{Approach to GPGPU parallelization}
The simulation of dynamical models in ecology and evolution is inherently serial (as the state variables are often Markovian). Thus, parallelization in sPEGG occurs within a time-step of the model across individuals. For many modeling problems, the benefits of parallelization, particularly on GPGPUs, are often greatest when the number of calculations are large (e.g., \citealt{harish07}, \citealt{nyland07}, and \citealt{owens08}). Hence, the performance gains due to sPEGG are largest when calculations within each time-step can be carried out over a large number (typically $>$ 10$^6$) of individuals. This may present a potential problem for accelerating simulations that model the eco-evolutionary dynamics of a smaller number of individuals, as can often be the case in some systems (e.g., \citealt{pelletier09}). 

A single simulation run of a given model may only simulate a small number of individuals - for instance, on the order of hundreds to tens of thousands, which are well below abundances at which the benefits of parallelization on the GPGPU may be apparent. However, sPEGG's power lies in its ability to simulate a large number of individuals across several simulation runs simultaneously. Consider simulating a model of 1000 individuals across 10,000 different parameter combinations. This exercise requires performing calculations on 10$^7$ individuals, which is well within the range of the number of calculations whose performance can benefit from parallelization on a GPGPU. 

To accomplish this, sPEGG organizes individuals into discrete patches. The population of individuals within a given patch is referred to as a deme. Individuals can potentially migrate between patches, although the user can also prohibit migration to and from patches. In sPEGG, patches that are not linked by migration can therefore be thought of as representing independent, replicate simulations. For instance, in the example above, sPEGG will simulate a model of 10,000 patches consisting of 1000 individuals each, where the model's dynamics within each patch are governed by a unique parameter combination. Supplementary Figure S5 illustrates an example of how sPEGG uses patches to enable parallel calculations to be performed simultaneously for all individuals. A given patch can contain multiple species, and, between migration events (if they are allowed), individuals only interact with other individuals within the same patch. 

\subsection*{Customizing sPEGG}
As noted in the main text, the sPEGG code-base for simulating a species-specific model requires some level of user customization. Functions initializing the data, simulating mating and reproduction, and updating the trait values/phenotypes of individuals are completely general and easily customizable for researchers using models to address a variety of questions. sPEGG includes built-in alternative routines and classes enabling researchers to readily implement code for diverse systems, such as different mating systems and inheritance mechanisms (see the main text for details).

For situations that are harder to generalize, such as the rules describing how individual phenotypes are updated in response to heterospecific individuals during each time step, sPEGG provides methods, as well as readily usable classes, to facilitate the updating of phenotypes. The case studies illustrate the use of these classes for performing calculations between individuals of different species as well as for modeling resource dynamics and individual consumption behavior. Finally, the individual-based models used in the case studies can be explored further by merely changing the numerical values in the corresponding configuration (text) files.

\subsection*{Comparison Between Serial and Parallel Versions of the Case Studies}
To assess the performance advantages of using sPEGG in each of the case studies in the main text, each IBM was re-coded by hand using equivalent C++ classes, iterators and functions (invoking functions from the widely-used GNU scientific library - \citealt{gsl} - which relies on inherently serial algorithms) for the functions, classes, and model-specific code in sPEGG. We then applied optimization techniques, which are known to improve execution time in a serial context, to the resulting code base. To compare the serial and parallel versions of our models, we simulated our serial IBMs on a single 3.6 Ghz (Intel core i7 3820) CPU processing core and used one half of an NVIDIA GTX 690 GPU (restricting program access to 2GB of GPU RAM and 1536 CUDA cores) to assess the parallel version's performance. All optimizing transformations provided by the compilers (g++ and nvcc) were enabled. 


\bibliographystyle{cbe}	
\bibliography{myrefs11}